\documentclass{amsart}

\pdfoutput=1
\usepackage[pdftex]{graphics}
\usepackage{amsmath,amssymb,mathtools}
\usepackage{graphicx}
\usepackage{xcolor}
\usepackage{epsdice}
\usepackage{enumitem}

\graphicspath{{./Figures/}}

\def\CA{\mathcal{A}}
\def\CW{\mathcal{W}}

\def\a{\alpha}

\def\sm{\mathfrak{m}}

\def\a{\alpha}\def\b{\beta}

\newcommand{\Li}{\ensuremath{\mathrm{Li}}}

\def\half{\frac{1}{2}}

\newcommand{\nn}{\nonumber}

\begin{document}

\author[Dongmin Gang]{Dongmin Gang}
\address{(DG and MY) Kavli IPMU, University of Tokyo, Kashiwa, Chiba 277-8583, Japan}

\author[Mauricio Romo]{Mauricio Romo}
\address{(MR) School of Natural Sciences, Institute for Advanced Study, Princeton NJ 08540, USA}

\author[Masahito Yamazaki]{Masahito Yamazaki}

\title[All-Order Volume Conjecture for Closed 3-Manifolds]
{All-Order Volume Conjecture for Closed 3-Manifolds from Complex Chern-Simons Theory}

\begin{abstract}
We propose an extension of the recently-proposed volume conjecture
for closed hyperbolic 3-manifolds, to all orders in perturbative expansion.
We first derive formulas for the perturbative expansion of the partition function of
complex Chern-Simons theory around a hyperbolic flat connection,
which produces infinitely-many perturbative invariants of the closed oriented 3-manifold.
The conjecture is that this expansion coincides with the perturbative expansion of the
Witten-Reshetikhin-Turaev invariants at roots of unity $q=e^{2\pi i/r}$ with $r$ odd, in the limit $r\to \infty$.
We provide numerical evidence for our conjecture.
\end{abstract}


\maketitle
\tableofcontents

\section{Introduction and Summary}

The goal of this paper is two-fold:
\begin{enumerate}[label=\Alph*]
\item We derive a perturbative expansion \eqref{CS_expansion}, around a hyperbolic flat connection, for the partition function of $SL(2,\mathbb{C})$ Chern-Simons theory \cite{Witten:1989ip,Witten:2010cx} on a general \emph{closed} hyperbolic oriented 3-manifold. Our starting point is the finite-dimensional integral expression \eqref{closed integral} for the partition function.

\item Based on the perturbative expansion  mentioned above, we present a new all-order perturbative extension \eqref{Generalized VC for closed} of the
recently-proposed volume conjecture \cite{ChenYang}
for Witten-Reshetikhin-Turaev (WRT) invariants \cite{Witten:1988hf,ReshetikhinTuraev} for closed oriented 3-manifolds.
\end{enumerate}
In the rest of this introduction let us explain these two points in more detail.


\subsection{Volume Conjecture for Knot Complements}

The celebrated volume conjecture \cite{KashaevLinkInvariant,MurakamiMurakami,MMOTY} (see \cite{Murakami_Introduction} for review) states a surprising relation
between two a-priori very different objects.

The first is the colored Jones polynomial \cite{JonesPolynomial} of a link (or a knot)\footnote{A link in general has several components. A link is called a knot if it has only one component.} $L$ in the three-sphere $S^3$. We denote this by $J_N(L; q)$,
where $N$ denotes the coloring (an integer specifying a representation of $SU(2)$),
and $q$ is a formal parameter.\footnote{The Jones polynomial referred here is the normalized Jones polynomial, namely an unknot has a trivial Jones polynomial. Also, the colored Jones polynomial in our convention is
a Laurent polynomial in $q$, and not in $q^{1/2}$. In the literature $q$ here is sometimes
denoted by $q^{1/2}$.}

The second is the complex hyperbolic volume\footnote{More generally this is a simplicial volume (Gromov norm) \cite{GromovVolume}. When $L$ is a hyperbolic link, namely when $S^3 \setminus L$ is hyperbolic, then the simplicial volume coincides with the hyperbolic volume. Since we are mostly interested in the
hyperbolic cases, we will hereafter mostly refer to this quantity as ``hyperbolic volume''.}  of the knot complement
$S^3\setminus L$, which is a combination of the hyperbolic volume together with the Chern-Simons invariant:
$\textrm{Vol}(L)+ i \textrm{CS}(L)$.

The volume conjecture \cite{KashaevLinkInvariant,MurakamiMurakami,MMOTY} states that the
asymptotic behavior of the root-of-unity value of the
former gives the latter:
\begin{align}
\label{VC_knot}
\lim_{N\to\infty} J_N(L; q=e^{\frac{\pi i}{N}}) =
\exp\left[
\frac{N}{2\pi}
\left(
\textrm{Vol}(L)+ i \textrm{CS}(L)
\right)
\right]
\;.
\end{align}
%

\subsection{Volume Conjecture for WRT Invariants}

A natural extension of the volume conjecture is to consider
a closed hyperbolic oriented 3-manifold $M$, where the relevant quantity replacing the Jones polynomial would be the
WRT invariant, which was formulated mathematically by Reshetikhin and Turaev \cite{ReshetikhinTuraev}
based on the physics idea of Witten  \cite{Witten:1988hf}.
This invariant is defined from a modular Hopf algebra,
which can be obtained from a quantum group $U_q(\mathfrak{sl}_2)$ when $q$ is a primitive root of unity.\footnote{Contrary
to the case of  the colored Jones polynomial where $q$ is a formal parameter,
WRT invariant is defined only for $q$ being a primitive root of unity.
If we try to construct and expression in $q$ whose values at root of unity reproduce the WRT invariants,
we need to consider a certain cyclotomic completion of the
polynomial ring in $q$ \cite{Habiro}.}
For $q=\exp(\pi i/r)$ we denote the associated invariant by $\tau^{\rm SU(2)}_{r}(M)$, where $r$ is an integer $r\ge 3$.\footnote{In the language of $SU(2)$ Chern-Simons theory this integer $r$ is the 1-loop corrected level $r=k+2$, where $k$ is an integer known as the level and is a parameter in front of the classical Chern-Simons action.}
It is then natural to consider the limit $r\to \infty$, and expect that we again reproduce the complex simplicial volume of the closed 3-manifold:
\begin{align}
\label{VC_incorrect}
\lim_{r\to\infty} \tau^{\rm SU(2)}_r(M) \overset{??}{=}
\exp\left[
\frac{r}{2\pi}
\left(\textrm{Vol}(M)+ i \textrm{CS}(M) \right)
\right]
\;.
\end{align}
It turns out, however, this naive conjecture does not work;
WRT partition function grows in a power-law in $r$, and not exponentially.\footnote{See nevertheless \cite{Murakami_WRT,Constantino,ConstantinoMurakami}
for related discussion, which applies a formal saddle point analysis to the WRT invariant $\tau_r^{\rm SU(2)}$ and obtained the complex volume.}

A new insight was brought by the recent work of
Chen and Yang \cite{ChenYang}, who considered the root-of-unity value $q=e^{2\pi i /r}$ with odd $r$.
Let us denote the associated WRT invariant by $\tau^{\rm SO(3)}_r(M)$, since this
invariant is known to be related to the group $SO(3)$, rather than $SU(2)$.
We will call this invariant the $SO(3)$-WRT invariant.
This invariant is
discussed for example by Kirby and Melvin \cite{KirbyMelvin},
and studied further by Blanchet et al.\ \cite{BHMV} and Lickorish \cite{Lickorish4} in the
context of skein-theory reformulation of WRT invariants by Lickorish \cite{Lickorish1,Lickorish2,Lickorish3}.\footnote{
$\tau^{\rm SO(3)}_r(M)$ ($r$ odd) was denoted by $\tau'$ in  \cite{KirbyMelvin}, and $\theta_r$ in \cite{BHMV}; they all coincide, up to overall normalization factors \cite{BHMV2}. Our normalization follows \cite{Lickorish4}.}

Now the conjecture due to \cite{ChenYang}
states the following asymptotics:
\begin{align}
\Aboxed{
\lim_{r \in 2\mathbb{Z}+1; r \rightarrow \infty}\tau^{\rm SO(3)}_{r}(M)
=
\exp\left[
\frac{r}{2\pi}
\left(
\textrm{Vol}(M)+ i \textrm{CS}(M)
\right)
\right]
} \;.
\label{VC for closed}
\end{align}

\subsection{All-Order Generalization}

We are now ready to state the main results of this paper.

First, we introduce a perturbative expansion for the state-integral model for complex Chern-Simons theory on a closed hyperbolic oriented 3-manifold $M$,
whose partition function (which we denote by $Z_{\hbar}(M)$) is written as a finite-dimensional integral.
We then consider the $\hbar$-expansion of this quantity around the complete hyperbolic flat connection:
\begin{align}
 Z_{\hbar}(M)
  \overset{\hbar\to 0}{\longrightarrow} \exp\left[ \frac{1}\hbar S^{\rm hyp}_0 (M) + S^{\rm hyp}_1(M) + \hbar S^{\rm hyp}_2(M)+\ldots  \right]\;.
  \label{CS_expansion}
  \end{align}
Here each of the expansion coefficient $S^{\rm hyp}_n(M)$ is an interesting perturbative invariant of the closed 3-manifold $M$.
We will give a concrete prescription for computing these invariants.\footnote{We can also consider non-perturbative evaluation of the state-integral along a proper converging integration cycle. The non-perturbative partition function can be identified as a Borel resummation of the perturbative expansion \cite{Bae:2016jpi}. Thanks to the so-called 3d--3d correspondence \cite{Terashima:2011qi,Dimofte:2011ju}, the partition function can  be interpreted as a partition function of a three-dimensional $\mathcal{N}=2$ supersymmetric gauge theory on  a curved background.
}

We then claim that the expansion \eqref{CS_expansion}
coincides with the $r\to \infty$ of the WRT invariant $\tau^{\rm SO(3)}_{r}(M)$:\footnote{More precisely,
the perturbative invariants $S_{n=0,1,2}$ has ambiguities as stated in \eqref{ambiguity of invariants}, and the match is
meant to be modulo this ambiguity.}
\begin{align}
\Aboxed{   \tau^{\rm SO(3)}_{r}(M)
  \overset{r \in 2\mathbb{Z}+1 ;r\rightarrow \infty }{\xrightarrow{\hspace*{1cm}}}
  \exp\left[
  \frac{1}\hbar S^{\rm hyp}_0 (M) + S^{\rm hyp}_1(M) + \hbar S^{\rm hyp}_2(M)+\ldots
  \right]
    \Big|_{\hbar  = \frac{2\pi i }r} }\;. \label{Generalized VC for closed}
\end{align}
Since the first coefficient $S_0(M)$ is shown to coincide with the
complex volume $i\textrm{Vol}(M)- \textrm{CS}(M)$,
\eqref{Generalized VC for closed}
contains and generalizes the conjecture \eqref{VC for closed}.

\bigskip
The rest of the paper is organized as follows.
In section 2 we explain how to obtain the integral formula for the
partition function for the closed 3-manifold, as motivated from complex Chern-Simons theory.
In section 3 we discuss the perturbative expansion of the partition function around a hyperbolic flat connection.
In section 4 we check for some examples the generalized conjecture \eqref{Generalized VC for closed} numerically.
We also include appendices for review materials.

\noindent
\textit{Note added:} After submission of the manuscript to arXiv, we have been noticed of the preprint by Tomotada Ohtsuki \cite{T.Ohtsuki},
which also discusses asymptotic expansion of WRT invariants.

\section{State-Integral Model for Closed 3-Manifolds}

In this section we introduce a partition function $Z_{\hbar}(M)$ for a closed oriented hyperbolic 3-manifold $M$.
We then go on to discuss its perturbative expansion around a hyperbolic flat connection.

\subsection{Derivation}

Our discussion of closed 3-manifolds relies on the famous mathematical theorem by Lickorish and Wallace \cite{Lickorish,Wallace},
which states that any closed 3-manifold can be obtained by a Dehn filling of the complement $S^3\setminus L$
of a link $L$ inside $S^3$.

The task of deriving a partition function is therefore divided into two.
The first is to derive a formula for the partition function for a link complement $S^3\setminus L$.
The second is to study the effect of the Dehn filling on the partition function.

\subsubsection{State-Integral Model for Link Complements}

For a knot/link complement, there are several developed  state-integral models \cite{2007JGP,Dimofte:2011gm,Andersen:2011bt}. For our purpose, in particular, we will use the state-integral model developed in \cite{Dimofte:2011gm,Dimofte:2012qj} (see also \cite{ohtsuki2015kashaev,ohtsuki2016asymptotic,ohtsukiasymptotic,ohtsuki2017asymptotic}  for discussion of higher order terms for knot complements).
This result was motivated from complex Chern-Simons theory \cite{Witten:1989ip,Witten:2010cx}; in our context
this is natural since Jones polynomial is nothing but the vacuum expectation value of the
Wilson line in Chern-Simons theory \cite{Witten:1988hf} and an interpretation for the volume conjecture (for a link complement) is provided in \cite{Gukov:2003na}.

Given a hyperbolic knot/link complement, we can consider its regular ideal triangulation.
Let us denote the number of ideal tetrahedra by $k$.
The gluing rules of the ideal tetrahedra are specified by the gluing datum $\{A, B, C, D, \vec{\nu}, \vec{\nu}' \}$.
Here $A, B, C, D$ are $k\times k$ matrices, and $\vec{\nu}, \vec{\nu}'$ are $k$-vectors.
For details, see Appendix \ref{sec.state_integral}.
 Then the state-integral expression for the link complement is \cite{Dimofte:2011gm,Dimofte:2012qj}
\begin{align}
\begin{split}
&
Z_{\hbar}^{ (\mathbb{X}_\a,\mathbb{P}_\a) } ( \hat{M}\setminus L; X_\a)
\\
&
= \frac{2}{\sqrt{\det B}}\int \prod_{i=1}^k \frac{dZ_i}{\sqrt{2 \pi \hbar}}  \exp\left( \frac{1}{\hbar} Q(\vec{Z}, \vec{X};  \{A,B,C,D,\vec{\nu}, \vec{\nu}'\} ) \right) \prod_{i=1}^k  \Psi_{\hbar} (Z_i) \;.\label{Dimofte integral}
\end{split}
\end{align}
This finite dimensional integration can be interpreted  as  a $SL(2,\mathbb{C})$ Chern-Simons partition function on the link complement $\hat{M}\backslash L$ with analytically-continuned Chern-Simons level $\frac{2\pi i}{\hbar}$.\footnote{It is known that this state-integral model does not capture reducible flat connections.
We will, however, be interested in the perturbative expansion of our partition function
around a hyperbolic flat connection, which is irreducible, and hence this subtlety is not important for the considerations of this paper.} %
Let us explain this formula in detail.
In order to compute the partition function
we need to specify the choice of polarization on the boundary of
$\hat{M}\setminus L$. In our case, this is to choose a basis of $H_1(\partial (\hat{M}\setminus L),\mathbb{Z})$,
and if the link $L$ has $S$ components we need to pick up $S$ pairs of generators
$(\mathbb{X}_\a,\mathbb{P}_\a)_{\a=1}^S$.
For our later purposes it is actually sufficient to restrict to the case $\hat{M}=S^3$.
Then we have a canonical choice
\begin{align}
\mathbb{X}_{\a}=\sm_{\a} \;(\textrm{merdian})\;, \quad \mathbb{P}_{\a} =\mathfrak{l}_{\a} \;(\textrm{longitude})\;.
\label{m_and_l}
\end{align}
for each boundary torus labeled by $\alpha$.
Once we fix a polarization $(\mathbb{X}_\a,\mathbb{P}_\a)_{\a=1}^S$ ,
the partition function depends non-trivially on
the deformation parameters (boundary conditions) $(X_{\alpha})_{\alpha=1}^S$;
their effect is to modify the holonomy along the meridian cycles of the knot complements
(see \eqref{eq.gluing_2} and \eqref{X_def} in appendix).
The integral \eqref{Dimofte integral} is over $k$ parameters $\{ Z_i\}_{i=1}^k$, one for each tetrahedron.
For the integrand, the expression $Q$ on the exponent is a
quadratic expression in $\vec{Z}$ and $\vec{X}$:\footnote{We did not include
$\vec{f}, \vec{f}''$ in the arguments of $Q$
since they are determined by other arguments \eqref{nu_def}.}
\begin{align}
\begin{split}
&Q(\vec{Z}, \vec{X}; \{A,B,C,D,\vec{\nu}, \vec{\nu}'\} ):=\frac{1}{2 } \vec{Z} B^{-1} A \vec{Z} + 2 \vec{X}  D B^{-1} \vec{X}+  (2\pi i +\hbar) \vec{f} B^{-1} \vec{X}  \\
&\qquad \qquad+\half \left(i \pi + \frac{\hbar}{2}\right)^2 \vec{f} B^{-1} \vec{\nu} - \vec{Z} B^{-1} \left(\left(i \pi + \frac{\hbar}{2}\right) \vec{\nu}+2 \vec{X}\right) \;.
\end{split}
\end{align}
The rest of the integrand is a product of the
quantum dilogarithm function
$\Psi_{\hbar} (Z)$, which is defined as (for $\textrm{Re}(\hbar)<0$)  \cite{FaddeevKashaevQuantum}\footnote{This function has a symmetry under
$b\to 1/b$, where $b$ is a parameter related to $\hbar$ by $\hbar=2\pi i b^2$. This symmetry, however, is not important for a perturbative consideration of this paper.}
\begin{align} \label{Q.D.L}
&\Psi_{\hbar} (Z) := \prod_{r=1}^\infty \frac{1- (e^{\hbar})^r e^{-Z}} {1-\left(e^{-\frac{(2\pi i)^2}{\hbar}}\right)^{r-1} e^{-2\pi i \frac{Z}{\hbar}}}\;.
\end{align}
The parameter $\hbar$ is the expansion parameter of the partition function,
which is to be identified with the parameter of the same name in \eqref{Generalized VC for closed}.

Given a knot complement, the choice of an ideal triangulation,
as well as its gluing datum, is far from unique.
It can be shown, however, that the partition function
\eqref{Dimofte integral} is independent of such choices up to the following ambiguity \cite{Dimofte:2012qj}\footnote{The $\frac{\pi^{2}}{6}$ ambiguity at $\hbar^{-1}$ can actually be lifted to $2\pi^{2}$ \cite{2003math......7092N}.}:%
\begin{align}
\exp\left(
\frac{\pi^2}{6\hbar} \mathbb{Z}+ \frac{i\pi}{4} \mathbb{Z}+\frac{\hbar}{24} \mathbb{Z}
\right) \;.
\label{ambiguity}
\end{align}
In the following $\hbar$ will be taken to be pure imaginary (see \eqref{pure_imaginary}), in which case the ambiguity is only a phase factor.
%

\subsubsection{Dehn-Filling Formula}

The second ingredient is the Dehn-filling formula,
which specifies the change of the partition function
under the Dehn filling (a similar formula for compact-group Chern-Simons theory is well-known, see \cite{Witten:1988hf}).

Consider a 3-manifold obtained by performing $(p_{\alpha}, q_{\alpha})$-Dehn filling $(\alpha=1, \dots, S')$
for the knot complement $\hat{M}\setminus L$ along  $S'(\leq S)$ components of a link $L$ out of $S$ components.
We denote this manifold as
\begin{align}
(\hat{M}\setminus L)_{ \{p_\a \mathbb{X}_\a +q_\a \mathbb{P}_\a\}_{\a=1}^{S'}} \;.
\end{align}
where in the notation $p_\a \mathbb{X}_\a +q_\a \mathbb{P}_\a$ denotes the cycle of the boundary torus
which becomes contractible after Dehn filling (see \eqref{Dehn_filling} in appendix).
Let us first assume that $q_{\a}\ne 0$ for all $\a$.
Then our Dehn-filling formula is given by \cite{Bae:2016jpi}\footnote{See also \cite{Alday:2017yxk} for recent discussion on Dehn fillings.}
\begin{align}
\begin{split}
& Z_\hbar^{(\mathbb{X}_\a, \mathbb{P}_\a)_{\a =S'+1}^S} \big{(} ( \hat{M}\setminus L)_{ \{p_\a \mathbb{X}_\a+q_\a \mathbb{P}_\a \}_{\a=1}^{S'}}; \{ X_{\a} \}_{\a = S'+1}^S\big{)}
\\
& \quad
=\int  \left( \prod_{\a=1}^{S'}\frac{dX_\a   }{\sqrt{2\pi   \hbar \, q_\a}}
\mathcal{K}_{p_{\a},q_{\a}}(X_\a) \right)
Z_\hbar^{(\mathbb{X}_\a, \mathbb{P}_\a)} ( \hat{M}\setminus L; X_\a )\;,
\label{Dehn_filling_formula}
\end{split}
\end{align}
where $s_\a $ (and $r_\a$) is defined by the condition\footnote{As discussed before $s_\a$ is defined up to $q_\a\mathbb{Z}$. This ambiguity does not change the formula \eqref{Dehn_filling_formula} modulo \eqref{ambiguity}. The formula is also invariant under the sign flip of $(p,q)$, which can be seen explicitly by noting that $Z_\hbar^{(\mathbb{X}_\a, \mathbb{P}_\a)} ( \hat{M}\setminus L; X_\a )=Z_\hbar^{(\mathbb{X}_\a, \mathbb{P}_\a)} ( \hat{M}\setminus L; -X_\a )$ by Weyl invariance and $\mathcal{K}_{p_{\a},q_{\a}}(X_\a)=\mathcal{K}_{-p_{\a},-q_{\a}}(-X_\a)$.}
\begin{align}
&  \left(\begin{array}{cc}r_\a & s_\a \\ p_\a & q_\a  \end{array}\right) \in PSL(2,\mathbb{Z}) \;,
\end{align}
and the integral kernel for the Dehn filling is given by
\begin{align}
\begin{split}
 \mathcal{K}_{p,q}(X):=&
\exp\left[\frac{s}{q} \left(\frac{\pi ^2 }{\hbar }-\frac{\hbar }{4 }\right) +\frac{p X^2}{ q \hbar}\right] \\
&\times \left(e^{\frac{2\pi i X}{\hbar q}} \sinh\left(\frac{X-i \pi s}q\right) -e^{-\frac{2\pi i X}{\hbar q}}\sinh\left(\frac{X+i \pi s}q\right)\right) \;.%
\end{split}
\end{align}
When $q_\a = 0$, the formula \eqref{Dehn_filling_formula} should be modified as
\begin{align}
\begin{split}
\int \frac{dX_\a   }{\sqrt{2\pi   \hbar \, q_\a}}
\mathcal{K}_{p_{\a},q_{\a}}(X_\a) \overset{q_\a=0}{\xrightarrow{\hspace*{1cm}}}  \int \frac{dY_\a dX_\a}{\sqrt{2}\pi \hbar} \sinh(Y_\a)\sinh\left(\frac{2\pi i Y_\a}{\hbar}\right)e^{- \frac{2X_\a Y_\a}{\hbar}}\;. \label{qa=0 case}
\end{split}
\end{align}
The resulting 3-manifold   has  $(S-S')$ cusp boundaries and its partition function $Z_{\hbar}$ depends on the same number of variables $\{X_\a \}_{\a=S+1}^S$. When $S=S'$, the 3-manifold $(\hat{M}\setminus L)_{ \{p_\a \mathbb{X}_\a +q_\a \mathbb{P}_\a\}_{\a=1}^{S}}$ is a closed 3-manifold.

It is worth pointing out that a version of the Dehn-filling formula was proposed in \cite[section 3]{2007JGP}. The integral kernel there coincides with the
leading semiclassical piece of our integral kernel $\mathcal{K}_{p,q}$,
\begin{align}
\mathcal{K}_{p,q}\overset{\hbar  \to 0}{\xrightarrow{\hspace*{1cm}}} \frac{1}{2} \exp\left[ \frac{1}{\hbar} \left(
\frac{p}{q}X^2 +\frac{2\pi i}{q} X+ \frac{\pi^2 s}{q}
\right)+\mathcal{O}(\hbar^0)%
\right] \;, \label{Hikami's model}
\end{align}
where we assumed $\mathrm{Re}(X/q)>0$.
This means that the two proposals give the same result as far as the
leading classical results are concerned.
Leading classical part of the state-integral model in \cite{2007JGP} is shown to give the complex hyperbolic volume of $M$,
so does our state-integral model.
However, the two proposals give
different answers for higher orders in $\hbar$, and the difference will crucially affect  the discussion of the volume conjecture below.

\subsection{Main Formula}

We are now ready to give the final expression for the state-integral model.
Suppose that a closed 3-manifold $M$ is obtained from the
knot complement $\hat{M}\setminus L$
by $(p_\a, q_\a)$-Dehn surgeries on the $\a$-th component: in our previous notation we have
$M=(\hat{M}\setminus L)_{\{p_\a \mathbb{X}_\a+q_\a\mathbb{P}_\a\}_{\a=1}^S}$.
For this 3-manifold, our formula is given by a concrete finite-dimensional integral expression\footnote{Our partition function
\eqref{closed integral} can be thought of as
finite-dimensional counterparts of the infinite-dimensional path integral.
Since the complex Chern-Simons theory has a complex gauge field,
its action is also complex and the precise definition of the path integral requires
a subtle choice of the integration contour \cite{Witten:2010cx}.
This is reflected in the choice of integration contour in \eqref{closed integral}.
These subtleties, however, are irrelevant for perturbative expansions discussed in this paper.},
which is obtained by \eqref{Dimofte integral} and \eqref{Dehn_filling_formula}:\footnote{We can also apply the Dehn filling prescription \eqref{Dehn_filling_formula} to the cluster partition of \cite{Terashima:2013fg,Gang:2015bwa,Gang:2015wya}. It would be interesting to study the resulting partition function.}
\begin{align}
\begin{split}
Z_{\hbar} (M)
&= \frac{2}{\sqrt{\det B}}\int  \prod_{\a=1}^S \frac{\mathcal{K}_{p_\a,q_\a}(X) dX_\a}{\sqrt{2\pi \hbar q_\a}} \prod_{i=1}^k \frac{\Psi_{\hbar} (Z_i)  dZ_i}{\sqrt{2 \pi \hbar}} \\
& \qquad \qquad \times
 \exp\left( \frac{1}{\hbar} Q(\vec{Z}, \vec{X};  \{A,B,C,D,\vec{\nu}, \vec{\nu}'\} ) \right) \;.\label{closed integral}
\end{split}
\end{align}
The formula is only valid for $q_\a\neq 0$ and it should be modified as \eqref{qa=0 case} when $q_\a=0$.
The expression of state-integral depends both on basis choice $(\mathbb{X}_\a, \mathbb{P}_\a)_{\a=1}^S$ of $H_1 \big{(}\partial (\hat{M}\backslash L),\mathbb{Z} \big{)}$ and filling slopes $(p_\a,q_\a)_{\a=1}^{S}$. The final state-integral, however, turns out to  depend only on the combination $\{ p_\a \mathbb{X}_\a+q_\a \mathbb{P}_\a \}_{\a=1}^{S}$ and invariant under following transformation:
\begin{align}
\begin{split}
&\left(\begin{array}{cc} \tilde{\mathbb{X}}_\a \\  \tilde{\mathbb{P}}_\a \end{array}\right) = \left(\begin{array}{cc}a_\a & b_\a \\ c_\a& d_\a \end{array}\right) \left(\begin{array}{cc} \mathbb{X}_\a \\  \mathbb{P}_\a \end{array}\right) \;, \qquad\left(\begin{array}{c} \tilde{p}_\a \\ \tilde{q}_\a \end{array}\right) = \left(\begin{array}{cc}d_\a & -c_\a \\ -b_\a & a_\a\end{array}\right) \left(\begin{array}{c}p_\a \\ q_\a \end{array}\right) \;,
\label{change_1}
\\
& \textrm{with } \left(\begin{array}{cc}a_\a & b_\a \\ c_\a & d_\a \end{array}\right) \in SL(2,\mathbb{Z})\;.
\end{split}
\end{align}
This is consistent with the obvious fact that the closed 3-manifold $M$ is invariant under the transformation.
The gluing datum $(A,B,\vec{\nu})$ and $(C,D, \vec{\nu}')$ depends on the choice of $\mathbb{X}_\a$ and $\mathbb{P}_\a$ respectively and let denote them as $(A_{\mathbb{X}},B_{\mathbb{X}},\vec{\nu}_{\mathbb{X}})$ and $(C_{\mathbb{P}}, D_{\mathbb{P}},\vec{\nu}'_{\mathbb{P}})$.
Under the transformation of $(\mathbb{X}_\a,\mathbb{P}_\a)$, the matrices transforms as
\begin{align}
\begin{split}
&\left(\begin{array}{ccc}(A_{\tilde{\mathbb{X}}})_{\a i} & (B_{\tilde{\mathbb{X}}})_{\a i} & (\nu_{\tilde{\mathbb{X}}})_\a\\ (C_{\tilde{\mathbb{P}}})_{\a i} & (D_{\tilde{\mathbb{P}}})_{\a i} & (\nu'_{\tilde{\mathbb{P}}})_\a \end{array}\right) = \left(\begin{array}{cc}a_\a & 2b_\a  \\ c_\a /2& d_\a \end{array}\right)  \left(\begin{array}{ccc}(A_\mathbb{X})_{\a i} & (B_{\mathbb{X}})_{\a i} & (\nu_{\mathbb{X}})_\a \\ (C_{\mathbb{P}})_{\a i} & (D_{\mathbb{P}})_{\a i}  & (\nu'_{\mathbb{P}})_\a  \end{array}\right) \;
\\
&\label{change_2}
\end{split}
\end{align}
for $\a=1,\ldots, S$  and other components do not transform.

Notice that the Dehn-filling representation of a closed 3-manifold is far from unique, and there are ambiguities associated with the Kirby moves \cite{KirbyCalculus}. While the connection with complex Chern-Simons theory suggest that our
partition function is invariant under such moves (up to possibly overall ambiguities discussed in \eqref{ambiguity}),
it would be interesting to prove this more directly from the formulas above. We leave the detailed proof for future work.

\section{Perturbative Expansion}

In this section, we work out the perturbative expansion of the partition function of the state-integral model \eqref{closed integral} for a
closed hyperbolic 3-manifold $M$.  For the expansion, we always assume that
\begin{align}
\hbar =2\pi i b^2  \in i \mathbb{R}_+ \;. \label{pure_imaginary}
\end{align}
Using the symmetry in \eqref{change_1} we can always make %
\begin{align}
q_\a \neq 0\;
\end{align}
and we  assume it  in the most  discussions of this section.
For the expansion, let us first expand the integrand of the state-integral model in powers of $\hbar$:
\begin{align}
\int  \prod_{\a=1}^S dX_\a\prod_{i=1}^k dZ_i  \exp \left( \sum_{n=0}^\infty \CW_{n} (\vec{Z}, X)\hbar^{n-1} \right)\;.
\label{expand}
\end{align}
For this expansion we can use the
$\hbar$-expansion of the quantum dilogarithm function $\Psi_{\hbar}$ \eqref{Q.D.L} \cite{Dimofte:2009yn}:
\begin{align}
\log \Psi_{\hbar} (Z) \sim \sum_{n=0}^\infty \frac{B_n \hbar^{n-1}}{n!} \Li_{2-n}(e^{-Z})\;, \quad \textrm{ for  $0< \textrm{Im}[Z] <\pi$\;. }
\end{align}
where $B_n$ is the $n$-th Bernoulli-Seki number with $B_1=1/2$.

By evaluating \eqref{expand} in the saddle point approximation (around a saddle point, whose choice we will discuss momentarily), we will obtain an expansion of the form \eqref{CS_expansion}.
Each of the $S_n(M)$ is a well-defined perturbative invariant of 3-manifolds,
with curious number-theoretic properties.
For $SU(2)$ Chern-Simons theory, such an expansion is considered for example in
\cite{AxelrodSinger_I,AxelrodSinger_II}. For $SL(2,\mathbb{C})$ Chern-Simons theory,
see \cite{Dimofte:2009yn,Dimofte:2012qj} for similar perturbative invariants for knot complements.

Due the ambiguity in \eqref{ambiguity}, the perturbative series $\{S_n(M)\}$ is defined up to
\begin{align}
S_0 \sim S_0 + \frac{\pi^2}6 \mathbb{Z}\;, \quad S_1 \sim S_1 +i \frac{\pi}{4} \mathbb{Z}\;, \quad S_2 \sim S_2 +\frac{\mathbb{Z}}{24}\;. \label{ambiguity of invariants}
\end{align}
Under the change of  orientation $M\rightarrow \overline{M}$,
\begin{align}
Z_{\hbar}(\overline{M}) = (Z_{\hbar}(M))^* \quad \Rightarrow \quad S_n(\overline{M}) = (-1)^{n+1}(S_n(M))^*\;.\label{orientation}
\end{align}

In the rest of this section we assume for simplicity of notation that $S=S'=1$;
namely, the closed 3-manifold $M$ is obtained by a Dehn filling along a one-component link (knot) $L$,
which we also denote by $K$, to match standard notation:
$M=(\hat{M}\setminus K)_{p\mathbb{X}+q\mathbb{P}}$.
It is straightforward to repeat the discussion for general values of $S$.

\subsection{Classical Part:  Complex Hyperbolic Volume} \label{sec.volume}

As already mentioned above,
to define the perturbative expansion $S^{(c)}_n$, we need to specify a saddle point $(X^{(c)},Z^{(c)})$.
For the formulation of the generalized volume conjecture for WRT invariants \eqref{Generalized VC for closed},
we are particularly interested in perturbative expansion $S_n^{\rm hyp}$ around saddle points corresponding to the hyperbolic structure of $M$.

To discuss this more explicitly, let us start wth the leading term of the integrand \eqref{expand}. For
which  is  given by ($\varepsilon :=\textrm{Sign}( \textrm{Re}[X/q])$)
\begin{align}
\begin{split}
& \CW_0 (X,\vec{Z})=  \CW_0 (\vec{Z})+  \frac{p}{q} X^2+ \varepsilon \frac{2\pi i }{q} X +\frac{\delta}{q} \pi^2 \\
&\qquad + \bigg{(} 2 \vec{X}  D B^{-1} \vec{X}+  (2\pi i ) f B^{-1} \vec{X} -\half \pi^2 \vec{f} B^{-1} \vec{\nu} - \vec{Z} B^{-1} \big{(}i \pi  \vec{\nu}+2 \vec{X}\big{)} \bigg{)} \;,
\\
 &\textrm{with }\CW_0 (\vec{Z})= \frac{1}{2} \vec{Z} B^{-1} A \vec{Z} + \sum_{i=1}^k \Li_2 (e^{-Z_i})
 \label{WZ} \;.
 \end{split}
\end{align}
Here the  vector $\vec{X}$ is $(X,0,\ldots,0)$ as defined in  eq.~\eqref{X_def}.
The formula \eqref{WZ} coincides with the so-called Neumann-Zagier potential for a knot complement \cite{NeumannZagier},
and the formula \eqref{WZ} recovers its transformation rule under the Dehn filling. Indeed, saddle point equations with respect to $Z_i$ are
\begin{align}
&\frac{\partial \CW_0}{ \partial Z_i} =  0 \quad \Rightarrow \quad  A \vec{ Z} +B  \vec{Z}'' = i \pi \vec{\nu}+ 2 \vec{X}\;,
\label{eom 1}
\end{align}
where  we defined $Z''_i:=\log (1- \frac{1}{z_i})$.
This equation coincides with the gluing equation \eqref{eq.gluing_1} upon exponentiation
and under the identification $e^{Z_i}=z_i, e^{Z''_i}=z''_i=1-z_i^{-1}$ (see also \eqref{zzbar}).
We also need to take into account
the saddle point equation for $X$:
\begin{align}
&\frac{\partial \CW_0}{ \partial X}
 =  0 \quad \Rightarrow \quad  p X + q P
 =   - \pi i \varepsilon  \;, \label{eom 3}
\end{align}
where
\begin{align}
P:= (C  \vec{Z}+ D  \vec{Z}'' - i \pi \vec{\nu}' )_{1}\;.  \label{eom 2}
\end{align}
In the above expressions, we simplify the equations using the equation of motion \eqref{eom 1} and the symplectic property of gluing matrices \eqref{Symplectic property}.
These saddle point equations \eqref{eom 1}, \eqref{eom 3}, \eqref{eom 2} are  equivalent to the gluing equations for closed 3-manifold studied by Neumann and Zagier \cite{NeumannZagier}.

Generically a solution to these equations gives a  $PSL(2,\mathbb{C})$ flat  connection on $M$. We can explicitly construct Hom$(\pi_1 (M)\rightarrow PSL(2, \mathbb{C}))$ from the solution \cite{Dimofte:2013iv,Gang:2015wya}. But there are  two subtle points as emphasized in \cite{Chung:2014qpa}. First, some flat-connections on a closed 3-manifold $M$, cannot be constructed from solutions of the gluing equations (\ref{eom 1}) and (\ref{eom 3}). This is because we are using a state-integral model based on ideal triangulations which do not capture reducible flat connections on a knot complement, $\hat{M}\backslash K$. This means flat connections on $M$ that originated from a reducible flat connection on the knot complement cannot be found as a saddle point in the state-integral model.  Second, some solutions of the gluing equations might not give a flat connection on $M$. The solutions of the gluing equations (\ref{eom 1}) and (\ref{eom 3}) are guaranteed to correspond to a flat connection on $\hat{M}\backslash K$ whose holonomy along $p\mathbb{X}+q\mathbb{P}$ has eigenvalue $\pm 1$. If the holonomy is $\pm$(identity), then it gives a $PSL(2,\mathbb{C})$ flat-connection on $M$. But if the holonomy is conjugate to $\pm$(parabolic)=$\pm \left(\begin{array}{cc}1 & 1 \\0 & 1\end{array}\right)$, it does not give a flat connection on $M$. In the latter case, the perturbative expansion around the parabolic solution cannot be interpreted as a perturbative expansion of $PSL(2, \mathbb{C})$ Chern-Simons theory on $M$. One simple example is $S^3 = (S^3\backslash K)_{\sm}$ with $K$ a hyperbolic knot. In this case, there are at least three flat-connections on $S^3\backslash K$ whose meridian ($\sm$) holonomy  has eigenvalue $\pm 1$. Two of them can be constructed from the unique complete hyperbolic structure on the knot complement, say $\mathcal{A}^{\rm hyp}$ and its conjugate $\mathcal{A}^{\overline{\rm hyp}}$.\footnote{A complete hyperbolic 3-manifold $M$ (knot complement or closed) with finite volume   can be realized as a quotient 3-dimensional hyperbolic upper half-plane, $\mathbb{H}^3/\Gamma$ with a discrete, torsion-free action $\Gamma$. The action $\Gamma$ gives a representation $\textrm{Hom}\big{(}\pi_1(M)\rightarrow PSL(2,\mathbb{C})=\textrm{Isom}^+(\mathbb{H}^3)\big{)}$ which defines a $PSL(2,\mathbb{C})$ flat-connection $\CA^{\rm hyp}$. Its conjugate representation defines $\CA^{ \overline{\rm hyp}}$.
In the language of three-dimensional gravity,  the flat connection $\CA^{\rm hyp}$ on hyperbolic 3-manifold  can be written as $\CA^{\rm hyp}=\omega + i e$ where $\omega$ and $e$ are spin connection and dreibein on $M$ constructed using the unique complete hyperbolic structure.  The conjugate flat connection is  $\CA^{\overline{\rm hyp}}= \omega - i e$.}
The third flat connection is an Abelian flat connection. General Abelian flat connection has trivial longitude holonomy and its meridian holonomy can be an arbitrary element of $PSL(2,\mathbb{C})$. The unique trivial $PSL(2,\mathbb{C})$ flat connection on $S^3$ comes from this Abelian flat connection on the knot complement which cannot be captured by an ideal triangulation.  Therefore, for models based on ideal triangulations, the solutions to the gluing equations correspond to the flat-connections $\CA^{\rm hyp}$ or $\CA^{\overline{\rm hyp}}$ which have $\pm$(parabolic) as meridian holonomy  and thus do not give a flat connection on $S^3$.

In the case that the resulting manifold $M$, after Dehn filling, is hyperbolic, the gluing equations (\ref{eom 1}) and (\ref{eom 3}) have a solution $\CA^{\overline{\rm hyp}}$ corresponding to the conjugate of the complete hyperbolic metric on $M$, satisfying
\begin{align}
0<\textrm{Im}[Z_i] <\pi \quad \textrm{for all $i=1,\ldots k$}\;.
\end{align}
This flat connection $\CA^{\overline{\rm hyp}}$ has the maximal value ($2 \textrm{Vol}(M)$) of $\textrm{Im}[\textrm{CS}[\CA]]$ among all $PSL(2, \mathbb{C})$ flat connections $\CA$ on $M$, where $\textrm{CS}[\CA]$ is the holomorphic Chern-Simons functional defined by
\begin{align}
\textrm{CS}[\CA]:= \int_M \textrm{Tr}\left[\CA \wedge d \CA + \frac{2}3 \CA^3\right]\;.
\end{align}
We can also consider perturbative series $\{S^{\rm hyp}_n \}$ around a flat connection $\CA^{\rm hyp}$ which has lowest value ($-2 \textrm{Vol}(M)$) of $\textrm{Im}[\textrm{CS}[\CA]]$. The two expansions are  related by complex conjugation,
\begin{align}
S^{\rm hyp}_n = (S^{\overline{\rm hyp}}_n)^* \quad \textrm{for all $n\geq 0$}\;. \label{relation between hyp and its conjugations}
\end{align}
Coming back to the flat connection $\CA^{\overline{\rm hyp}}$,
there is actually a two-fold degeneracy as originated from the
$\mathbb{Z}_2$ Weyl-symmetry of $SL(2)$; the two saddle points are
$(X^{\overline{\rm hyp},\varepsilon=1},Z^{\overline{\rm hyp},\varepsilon=1})$ and $(X^{\overline{\rm hyp},\varepsilon=-1},Z^{\overline{\rm hyp},\varepsilon=-1})$, with $X^{\overline{\rm hyp},\varepsilon=1}=-X^{\overline{\rm hyp},\varepsilon=-1}$.
Perturbative expansions around two saddle points are expected to be equal to all orders
\begin{align}
S^{\overline{\rm hyp},\varepsilon=1}_n = S^{\overline{\rm hyp},\varepsilon=-1}_n \quad \textrm{for all $n\geq 0$}\;.
\label{Weyl_reflection}
\end{align}
 due to the Weyl-symmetry. The leading classical contributions from the two saddle points $S^{\overline{\rm hyp}, \varepsilon}_n$
thus sum up to
\begin{align}
\begin{split}
&\exp\big{(}\sum_{n=0}^\infty S^{\overline{\rm hyp}}_n \hbar^{n-1}\big{)} = \exp\big{(}\sum_{n=0}^\infty S^{\overline{\rm hyp},\varepsilon=1}_n \hbar^{n-1}\big{)}
+\exp\big{(}\sum_{n=0}^\infty S^{\overline{\rm hyp},\varepsilon=-1}_n \hbar^{n-1}\big{)}
\\
&\quad \Longrightarrow  S^{\overline{\rm hyp}}_n =S^{\overline{\rm hyp},\varepsilon=1}_n+ \delta_{n,1}\log 2 \;.
\label{log2}
\end{split}
\end{align}
As stated above around \eqref{Hikami's model}, the classical part $S^{\overline{\rm hyp}}_0$   coincides with the complex hyperbolic volume of $M$:
\begin{align}
S^{\overline{\rm hyp}}_0 (M) =- i \textrm{Vol}(M) -\textrm{CS}(M) :=- \frac{1}2\textrm{CS}[\CA^{\overline{\rm hyp}}] \;.
\end{align}
%

\subsection{One-Loop Part: Reidemeister Torsion}\label{sec.1-loop}

Having specified the saddle point we can now discuss the perturbative expansion \eqref{CS_expansion}.
We define from the next $\mathcal{O}(\hbar^0)$ term in the expansion a quantity
\begin{align}
&\tau(M):=e^{-2 S^{\overline{\rm hyp}}_1 (M)} \;.
\end{align}
Then, as will be derived in  section~\ref{sec : higher order}
\begin{align}
\begin{split}
&\tau(M)= \pm \frac{1}8  \det (A \Delta_{z''} +B \Delta^{-1}_{z} - R \cdot \Delta_{z''}) \frac{p+2q (D B^{-1})_{1,1}}{( \sinh[\frac{X-\varepsilon i\pi s}q ])^2} \prod_{i=1}^k z_i^{f''_i} (z''_i)^{-f_i} \bigg{|}_{(z,X)\rightarrow (e^{Z},X)^{\overline{\rm hyp},\varepsilon}}\;,
\\
& \textrm{where }
R_{ij}:= \frac{2 q \; \delta_{i 1} (B^{-1})_{j,1}}{p +2 q (D B^{-1})_{1,1}}\;, \quad \Delta_{t}:=\textrm{diag}(t_1,\ldots, t_k)\quad t\in \{z,z',z'' \} \;.\label{torsion}
\end{split}
\end{align}
Using the equations of motion, the factor $\sinh[\frac{X- \varepsilon i \pi s}{q}]^2$ can be written in terms of edge variables
\begin{align}
\begin{split}
&\sinh\left[\frac{(A \vec{Z}+B \vec{Z}''-i \pi \nu)_1 - 2 \varepsilon i \pi s}{2q}\right]^2\; \quad \textrm{or equivalently,}
\\
&\sinh\left[\frac{(C \vec{Z}+D \vec{Z}''-i \pi \nu')_1 - i \varepsilon \pi r}{p}\right]^2\;.
\end{split}
\end{align}
In the derivation we have assumed $q$ is non-zero.
Nevertheless, the formula gives the correct answer even for $q=0$, if we use the second expression above.

There is a useful trick for the evaluation of $\tau(M)$.
Using the transformation \eqref{change_1} and \eqref{change_2}, we can map  $[(\mathbb{X},\mathbb{P});(p,q)]$ to $[(p\mathbb{X}+q \mathbb{P}, -r \mathbb{X}-s \mathbb{P});(1,0)]$. After transformation, the 1-loop is expressed as
\begin{align}
\begin{split}
\tau(M)&= \pm \frac{1}8  \det (A_{p\mathbb{X}+q\mathbb{P}} \Delta_{z''} +B_{p\mathbb{X}+q\mathbb{P}} \Delta_{z^{-1}} )\\
&\quad \frac{1}{( \sinh[rX+s P])^2} \prod_{i=1}^k z_i^{f''_i} (z''_i)^{-f_i} \bigg{|}_{(z,X)\rightarrow (e^{Z},X)^{\overline{\rm hyp},\varepsilon}}\;.
\label{1-loop}
\end{split}
\end{align}
Where $A_{p\mathbb{X}+q\mathbb{P}}$ and $B_{p\mathbb{X}+q\mathbb{P}}$ are the $A,B$ matrices corresponding to the polarization $p\mathbb{X}+q\mathbb{P}$. As checked in \cite{Dimofte:2012qj}, by overwhelming experimental evidence, is expected that
\begin{align}
\begin{split}
 &\textrm{Tor}(\hat{M}\setminus L; p\mathbb{X}+q\mathbb{P}) \\
 &\qquad =  \pm \frac{1}2  \det (A_{p\mathbb{X}+q\mathbb{P}} \Delta_{z''} +B_{p\mathbb{X}+q\mathbb{P}} \Delta_{z^{-1}} )\prod_{i=1}^k z_i^{f''_i} (z''_i)^{-f_i} \bigg{|}_{(z,X)\rightarrow (e^{Z},X)^{\overline{\rm hyp},\varepsilon}}\;, \label{Ray-singer torsion}
 \end{split}
\end{align}
where $\textrm{Tor}(\hat{M}\setminus L; p\mathbb{X}+q\mathbb{P}) $ denotes the Reidemeister torsion of adjoint representation twisted by the flat connection $\CA^{\overline{\rm hyp}}$ on $\hat{M}\setminus L$ associated  to the boundary one-cycle $p\mathbb{X}+q\mathbb{P}$.
This means that under Dehn filling
\begin{align}
\tau(M) = \textrm{Tor}(\hat{M}\setminus L; p\mathbb{X}+q\mathbb{P}) \frac{1}{4\sinh[rX+s P]^2}\;. \label{Ray-singer torsion 2}
\end{align}
This is exactly the same as the change of torsion under the Dehn filling (see, for example, \cite{2015arXiv151100400P})
\begin{align}
\textrm{Tor}\big{(}(\hat{M}\setminus L)_{p\mathbb{X}+q\mathbb{P}}\big{)} = \textrm{Tor}(\hat{M}\setminus L; p\mathbb{X}+q\mathbb{P}) \frac{1}{4\sinh[rX+s P]^2}\;. \label{Ray-singer torsion 3}
\end{align}
We have therefore proven (modulo the assumption of \eqref{Ray-singer torsion})
 that our 1-loop invariant coincides with the Reidemeister torsion:
\begin{align}
 \tau(M) = \textrm{Tor}(M)\;.
\end{align}
This is  an expected result since the 1-loop part of Chern-Simons partition function is given by the Reidemeister torsion \cite{Witten:1989ip}.

\subsection{Higher Order Results from Feynman Diagrams} \label{sec : higher order}

Let us now we will derive the Feynman rules, which are useful for
systematic computation of higher-order perturbative invariants.

Our starting point is the formula \eqref{Dehn_filling_formula}. We are interested in the limit $\hbar\rightarrow 0$.
As we already discussed in section \ref{sec.volume}, depending on the sign of $\mathrm{Re}(X/q)$, the dominant term in the exponential of the integral kernel $\mathcal{K}_{p,q}(X)$ will be either $\frac{p}{q}X^2\pm \frac{X}{q}$.
Under these assumptions, the leading pice of \eqref{Dehn_filling_formula}
in the limit $\hbar\rightarrow 0$ yields to the potential $\mathcal{W}_0(\vec{X},\vec{Z})$ \eqref{WZ}.
We have also seen in section \ref{sec.1-loop}
that there is a two-fold degeneracy in the saddle point corresponding to the sign $\varepsilon=+1$ and $\varepsilon=-1$,
with the same perturbative expansion to all orders \eqref{Dehn_filling_formula}.
This means we only need to focus on one choice of $\varepsilon$. We are then left with the following integral:
\begin{align}
\varepsilon e^{\frac{s}{q}\left(\frac{\pi^2}{\hbar}-\frac{\hbar}{4}\right)}\int \!\frac{dX}{\sqrt{2\pi \hbar q}}\, e^{\varepsilon\frac{2\pi i X}{q \hbar}+\frac{pX^{2}}{q\hbar}} \sinh \left(\frac{X-i\pi s \varepsilon}{q}\right)  Z_{\hbar}^{ (\mathbb{X},\mathbb{P}) } ( \hat{M}\setminus L; X)\;.
\end{align}
Let us combine all the integration variables $X$ and $(Z_i)_{i=1, \dots k}$ together into a $(k+1)$-component vector $y:=(y_a)_{a=1, \ldots, k+1}=(X,\vec{Z})$. Let us choose a critical point $y^{(c)}=(X^{c},\vec{Z}^{c})$ of $\mathcal{W}_0$.
The integral we want to perform perturbatively is given by:
\begin{align}\label{Zexp}
\begin{split}
Z^{(c),\varepsilon}_{\rm pert}(M)&=\frac{2\varepsilon}{\sqrt{\mathrm{det}B}}e^{\Gamma^{(0)}}\int [dy]\, e^{\frac{1}{2\hbar}H^{ab}y_{a}y_{b}}\sinh\left(\frac{X+X^{c}-i\pi s \varepsilon}{q}\right)
e^{(fB^{-1})_1 X}\prod_{i=1}^{k}e^{\sum_{s\geq 1}\Gamma_{i}^{(s)}/s!Z_{i}^{s}}  \\
&=\frac{2\varepsilon}{\sqrt{\mathrm{det}B}}e^{\Gamma^{(0)}} \int [dy]\, e^{\frac{1}{2\hbar}H^{ab}y_{a}y_{b}}\prod_{a=1}^{k+1}e^{\sum_{s\geq 1}\Gamma_{a}^{(s)}/s! y_{a}^{s}}
\;.
\end{split}
\end{align}
Here the measure is
\begin{eqnarray}
[dy]:=\frac{dX}{\sqrt{2\pi \hbar q}}\prod_{j=1}^{k}\frac{dZ_{j}}{\sqrt{2\pi \hbar }} \;,
\end{eqnarray}
In the first line \eqref{Zexp}
we have defined some symbols, including
the classical action:
\begin{align}
\begin{split}
\Gamma^{(0)}&=\frac{1}{\hbar}\mathcal{W}(y^{(c)})+\sum_{i}\sum_{n=1}^{\infty}B_{n}\frac{\hbar^{n-1}}{n!}\mathrm{Li}_{2-n}(e^{-Z^{c}_{i}})+(fB^{-1})_{1}X^{c} \\
&+\frac{i}{2}(\pi+\frac{\hbar}{4})fB^{-1}\nu-\frac{1}{2}Z^{c}B^{-1}\nu+\frac{s}{q}\left(\frac{\pi^2}{\hbar}-\frac{\hbar}{4}\right)
\;,
\end{split}
\end{align}
the linear vertex:
\begin{eqnarray}
\Gamma^{(1)}_{i}=-\frac{1}{2}(B^{-1}\nu)_{i}-\sum_{n=1}^{\infty}B_{n}\frac{\hbar^{n-1}}{n!}\mathrm{Li}_{1-n}(e^{-Z^{c}_{i}}) \;,
\end{eqnarray}
the quadratic vertex:
\begin{eqnarray}
\Gamma^{(2)}_{i}=\sum_{n=1}^{\infty}B_{n}\frac{\hbar^{n-1}}{n!}\mathrm{Li}_{-n}(e^{-Z^{c}_{i}}) \;,
\end{eqnarray}
the Hessian matrix:
\begin{align}
\begin{split}
&H^{ab}:=\frac{\partial \mathcal{W}(y^{(c)})}{\partial y_{a}\partial y_{b}} \;, \\
&H^{ij}=(B^{-1}A)_{i,j}+(z_{j}z''_{j})^{-1}\delta_{i,j} \;, \qquad H^{ix}=-2(B^{-1})_{i,1} \;,\qquad\\
& H^{xx}=2\frac{p}{q}+4(DB^{-1})_{1,1}
\;,
\end{split}
\end{align}
and finally the higher order vertices:
\begin{eqnarray}
\Gamma^{(s)}_{i}=(-1)^{s}\sum_{n=0}^{\infty}B_{n}\frac{\hbar^{n-1}}{n!}\mathrm{Li}_{2-n-s}(e^{-Z^{c}_{i}})\qquad s\geq 3 \;.
\label{higher_vertex}
\end{eqnarray}
In the second line \eqref{Zexp},
the sine-hyperbolic piece has been expanded as
\begin{align}
\sinh\left(\frac{X+X^{c}-i\pi s \varepsilon}{q}\right)=\sinh\left(\frac{X^{c}-i\pi s \varepsilon}{q}\right)\exp\left(\frac{X}{q}-\sum_{s=1}^{\infty}\frac{X^{s}}{s!}C_{s}(X^{c})\right) \;,
\label{sinh_term}
\end{align}
where
\begin{eqnarray}
C_{s}(X^{c}):=\left(-\frac{2}{q}\right)^{s}\mathrm{Li}_{1-s}\left(e^{-2\left(\frac{X^{c}-i\pi s \varepsilon}{q}\right)}\right) \;.
\end{eqnarray}
We can then define the combined $s$-vertex  $\Gamma^{(s)}_{a}$
for $y=(X, \vec{Z})$  by combining
\eqref{higher_vertex}, \eqref{sinh_term} and the exponential linear term in $X$ in \eqref{Zexp}:
\begin{eqnarray}\label{s-vertex}
\Aboxed{ \textrm{Vertices : }\Gamma^{(s)}_{a}:=\big{(} (fB^{-1})_1+q^{-1})\delta_{s,1}-C_{s}(X^{c}),\Gamma^{(s)}_{i} \big{)}} \;.
\end{eqnarray}

With this information, we can obtain the perturbative expansion of (\ref{Zexp}):
\begin{eqnarray}\label{pertexp}
Z^{(c),\varepsilon}_{\rm pert}(M)=\exp\left( \frac{1}{\hbar}S^{(c),\varepsilon}_{0}+\sum_{n=1}^{\infty}S^{(c),\varepsilon}_{n}\hbar^{n-1}\right) \;,
\label{CS_expansion_ep}
\end{eqnarray}
where the index $(c)$ in (\ref{pertexp}) is labelling the choice of critical point $y^{(c)}$.
The first two terms in the $\hbar$-expansion \eqref{CS_expansion_ep} are given by
\begin{align}\label{firstterms}
\begin{split}
S^{(c),\varepsilon}_{0}&=\mathrm{coeff}[\Gamma^{(0)},\hbar^{-1}] \;, \\
\exp(S_{1}^{(c),\varepsilon})&=2\varepsilon\sinh\left(\frac{X^{c}-i\pi s \varepsilon}{q}\right)\frac{i^{k+1}}{\sqrt{q\mathrm{det}(B)\mathrm{det}(H)}}e^{\mathrm{coeff}[\Gamma^{(0)},\hbar^{0}]} \;.
\end{split}
\end{align}
Here for a given a Laurent series $f(\hbar)$ on $\hbar$, $\mathrm{\mathrm{coeff}}[f(\hbar),\hbar^{a}]$ denotes
the coefficient of $\hbar^{a}$ in $f(\hbar)$.
We can verify that the expression (\ref{firstterms}) reduces to
\begin{align}\label{torr}
e^{S_{1}^{(c),\varepsilon}}=2\varepsilon\sinh\left(\frac{X^{c}-i\pi s \varepsilon}{q}\right)\frac{i^{k+1}\prod_{j}z_{j}^{-\frac{f''_{j}}{2}}(z''_{j})^{-\frac{f_{j}}{2}}}{\sqrt{(2p+4q(DB^{-1})_{1,1})\mathrm{det}(A\Delta_{z''}+B\Delta_{z^{-1}-R\Delta_{z''}})}} \;.
\end{align}
After including a factor $2$ from \eqref{log2} we can verify eq.~\eqref{torsion}.

The higher order terms in the $\hbar$ expansion can be computed by the Feynman diagram techniques. The situation is very analogous to \cite{Dimofte:2012qj}, except here we have the $\sinh$ term as in \eqref{sinh_term} and the vertex \eqref{s-vertex} is more involved.
The terms $S_{n> 1}$ will be extracted from a sum of connected graphs. Consider  a connected graph $\mathcal{G}_{\Gamma}$ with vertices of valences $k\geq 1$. Then, we associate a weight to $\mathcal{G}_{\Gamma}$: to each $k$-vertex we associate a factor $\Gamma^{(k)}_{t_{i}}$ and a label $t_{i}$ and to each internal line connecting two vertices with labels $t_{i}$ and $t_{j}$ a factor $\Pi_{t_{i},t_{j}}$, where we defined
the propagator:
\begin{align}
\Aboxed{\textrm{Propagator : }\Pi_{a,b}:=-\hbar(H^{-1})_{a,b}} \;,
\end{align}
Then the weight associated to the graph $\mathcal{G}_{\Gamma}$ is:
\begin{eqnarray}
W_{\Gamma}(\mathcal{G}_{\Gamma}):=\frac{1}{|\mathrm{Aut}(\mathcal{G}_{\Gamma})|}\sum_{\mathrm{labels}}\prod_{v\in \mathrm{vertices}}(-1)^{k_{v}}\Gamma^{(k_{v})}_{t_{v}}\prod_{e\in \mathrm{edges}}\Pi_{e} \;,
\end{eqnarray}
where $|\mathrm{Aut}(\mathcal{G}_{\Gamma})|$ is the symmetry factor (the rank of the group of automorphisms of $\mathcal{G}_{\Gamma}$). Given a connected graph $\mathcal{G}_{\Gamma}$ then is easy to see that $W_{\Gamma}(\mathcal{G}_{\Gamma})$ is of order $\hbar^{-V+E}$ or higher, where $E$ is the number of internal lines and $V$ the number of vertices with valence $k\geq 3$ in $\mathcal{G}_{\Gamma}$. After some computation one can show that $E=V+\mathcal{L}+V_{1}+V_{2}-1$ where $\mathcal{L}$ is the number of loops and $V_{1}$, $V_{2}$ are the number of $1$ and $2$-vertices respectively.

The Feynman rule for the perturbative invariant is then
\begin{eqnarray}
S^{(c),\varepsilon}_{n}=\mathrm{coeff}\left[\Gamma^{(0)}+\sum_{\mathcal{G}_{\Gamma}\in \mathcal{G}_{n}}W_{\Gamma}(\mathcal{G}_{\Gamma}),\hbar^{n-1}\right]\qquad n\geq 2 \;,
\end{eqnarray}
where we defined
\begin{eqnarray}
\mathcal{G}_{n}:=\{ \text{  Connected graphs \ }\mathcal{G}_{\Gamma}\text{ \ such that \ }\mathcal{L}+V_{1}+V_{2}\leq n \} \;.
\end{eqnarray}
For example, $S^{(c),\varepsilon}_{2}$ is given by:
\begin{align}
\begin{split}
&\frac{1}{4}\sum_{i}\mathrm{Li}_{0}\left(e^{-Z^{c}_{i}}\right)+\frac{i}{8}fB^{-1}\nu-\frac{s}{4q}+\mathrm{coeff}\Big[\frac{1}{8}\Gamma^{(4)}_{a}(\Pi_{aa})^{2}
+\frac{1}{8}\Pi_{aa}\Gamma^{(3)}_{a}\Pi_{ab}\Gamma^{(3)}_{b}\Pi_{bb}\\
&+\frac{1}{12}\Gamma^{(3)}_{a}(\Pi_{ab})^{3}\Gamma^{(3)}_{b}
+\frac{1}{2}\Gamma^{(1)}_{a}
\Pi_{ab}\Gamma^{(3)}_{b}\Pi_{bb}+\frac{1}{2}\Gamma^{(2)}_{a}\Pi_{aa}+\frac{1}{2}\Gamma^{(1)}_{a}\Pi_{ab}\Gamma^{(1)}_{b},\hbar\Big]  \;.
\end{split}
\end{align}

\subsection{Examples}

Let us discuss an example of $(S^3 \setminus \mathbf{4}_1)_{p\sm+q\mathfrak{l}}$. Here $\mathbf{4}_1$ denote the figure-eight knot, the simplest hyperbolic knot.
Since $\hat{M}=S^3$ in this case, we can use the canonical choices
\eqref{m_and_l}.
In this choice the gluing datum $(A,B,C,D, \vec{\nu},\vec{\nu}')$ for the figure-eight knot complement ($S^3\setminus \mathbf{4}_1$) are \cite{snappy}
\begin{align}
\begin{split}
&A = \left(\begin{array}{cc}1 & 0 \\ -1 & -1 \end{array}\right)\;, \quad B = \left(\begin{array}{cc}0 & -1 \\ 1 & 1 \end{array}\right)\;, \quad C= \left(\begin{array}{cc}-1 & 0 \\ 0 & 0 \end{array}\right) \;, \quad D= \left(\begin{array}{cc}1 & 0 \\ 0 & -1 \end{array}\right)
\\
& \vec{\nu} = \left(\begin{array}{cc}0  \\ 0 \end{array}\right)\;, \quad \vec{\nu}' = \left(\begin{array}{cc}0  \\ 0 \end{array}\right)\;.
\end{split}
\end{align}
Using the gluing datum and the perturbative expansion developed in previous section, we can compute $S^{\overline{\rm hyp}}_n \big{(}(S^3\setminus \mathbf{4}_1)_{p\sm+q \mathfrak{l}}\big{)}$. The knot is amphichiral  and thus topologically $(S^3\setminus \mathbf{4}_1)_{p \sm+q \mathfrak{l}}=(S^3\setminus \mathbf{4}_1)_{-p \sm+q\mathfrak{l}}$ for all $(p,q)$s.
For $M=(S^3\setminus \mathbf{4}_1)_{\pm 5 \sm+\mathfrak{l}}$, which is called Thurston manifold,
the saddle point (for $\varepsilon=+1$) is
\begin{align}
&(X,Z_1, Z_2)^{\overline{\rm hyp},\varepsilon=1} =(0.360784 - 0.575606 i, 1.59632 + 0.348931 i, -0.929172 + 1.23658 i)  \;,\nn
\end{align}
and the perturbative invariants are
\begin{align}
\begin{split}
&S^{\overline{\rm hyp}}_0 \big{(} M\big{)}=1.52067 - 0.981369 i  \;,
 \; S^{\overline{\rm hyp}}_1 \big{(} M\big{)} =-0.343697 + 3.78189 i \;,
\\
& S^{\overline{\rm hyp}}_2 \big{(} M\big{)} =-0.512461 + 0.0155225 i \;,  \; S^{\overline{\rm hyp}}_3 \big{(} M\big{)}=0.00927226 +0.00617571 i \;,
\\
& S^{\overline{\rm hyp}}_4 \big{(}M\big{)} =0.00312943 +0.00434039 i \;,  \; S^{\overline{\rm hyp}}_5 \big{(} M\big{)} =0.00164586 +0.00407186 i \;. \label{perturbative expansion for Thurston}
\end{split}
 \end{align}
We compute the perturbative invariants for $(p,q)=(5,1)$ and the invariants for $(p,q)=(-5,1)$ are simply related by the orientation change \eqref{orientation}. As another example, for $M=(S^3\setminus \mathbf{4}_1)_{- \sm+2\mathfrak{l}}$, we have
\begin{align}
\begin{split}
&S^{\overline{\rm hyp}}_0 \big{(}M\big{)} =  4.86783 - 1.39851 i\;, \;  S^{\overline{\rm hyp}}_1 \big{(}M \big{)}=-0.340874 - 3.9077 i \;,
\\
&S^{\overline{\rm hyp}}_2 \big{(}M \big{)} =  -0.610686 + 0.0259448 i \;, \;  S^{\overline{\rm hyp}}_3 \big{(}M \big{)}=0.0130034 + 0.00708517 i \label{perturbative expansion for (4-1)(-1,2)}\;.
\end{split}
\end{align}
%

\section{Numerical Evidence for All-Order Volume Conjecture}\label{sec.numerical}
Finally, let us present a numerical evidence for our conjecture \eqref{Generalized VC for closed} based on the technical developments in the previous sections.
The $SO(3)$-WRT invariant for the closed 3-manifold $(S^3\setminus  K)_{p\sm+\mathfrak{l}}$ is given
by \cite{Lickorish4}\footnote{This should be compared with
\begin{align}
\tau_r^{SU(2)} \big{(}(S^3\setminus  K)_{p\sm+\mathfrak{l}} \big{)} = \sqrt{\frac{2}r} \frac{1}{\sin(\frac{\pi}r)}\exp(\frac{(3-2p) \pi i}{4}) \sum_{N=1}^{r-1} \sin^2 (\frac{\pi N} r )  e^{ \frac{\pi i pN^2}2  }  J_{N}(K ;e^{\frac{\pi i }r }) \;. \nn
\end{align}}
\begin{align}
\begin{split}
&\tau^{SO(3)}_{r}\big{(}(S^3\setminus  K)_{p\sm+\mathfrak{l}} \big{)}
\\
&= \frac{2}r e^{\pi i (\frac{3+r^2}r- \frac{3-r}4)} \bigg{(} \sum_{N=0}^{r-2} \left(\sin \frac{2(N+1)\pi}r\right)^2 (-e^{\frac{\pi i}r})^{-p(N^2+2N)} J_{N+1}(K ;e^{\frac{2\pi i }r }) \bigg{)}\;. \label{so(3) WRT}
\end{split}
\end{align}
where $J_N (K;e^{\frac{2\pi i}r})$ is the value of $N$-th colored Jones polynomial of $K$ at $q=e^{\frac{2\pi i }r} $ with a normalization $J_N (\textrm{unknot})=1$.

\begin{figure}[htbp]
\begin{center}
   \includegraphics[width=.45\textwidth]{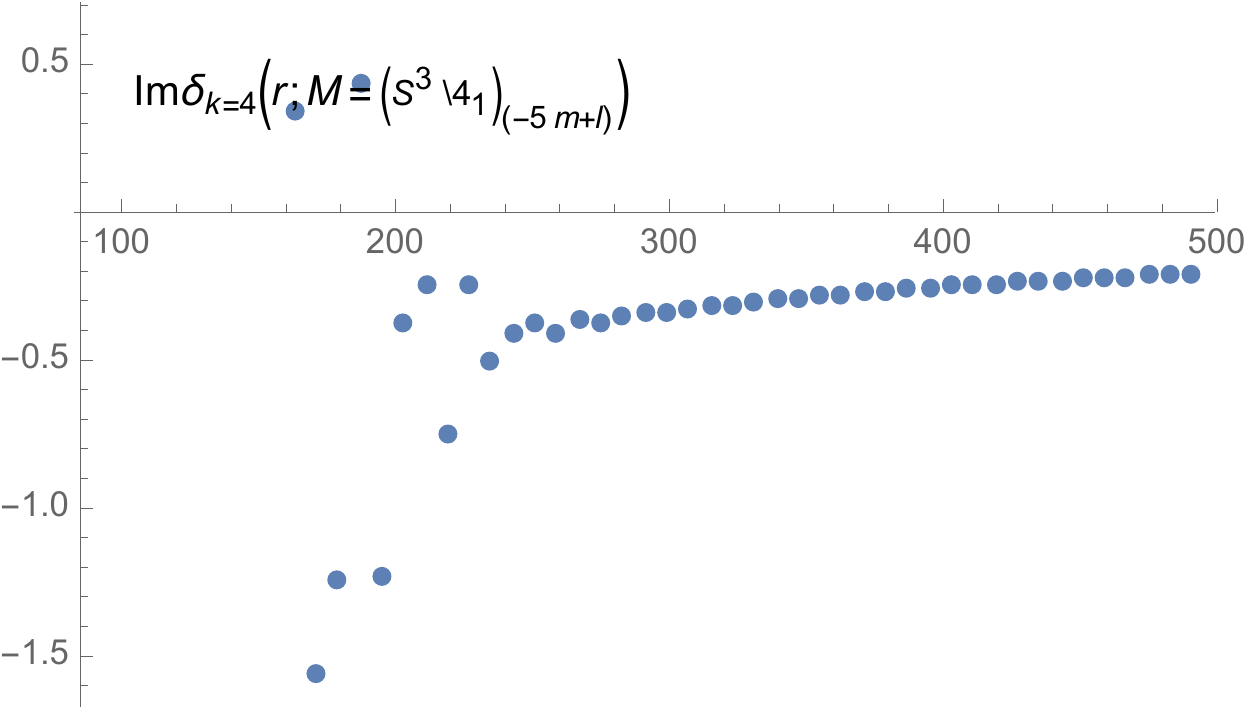}
    \includegraphics[width=.45\textwidth]{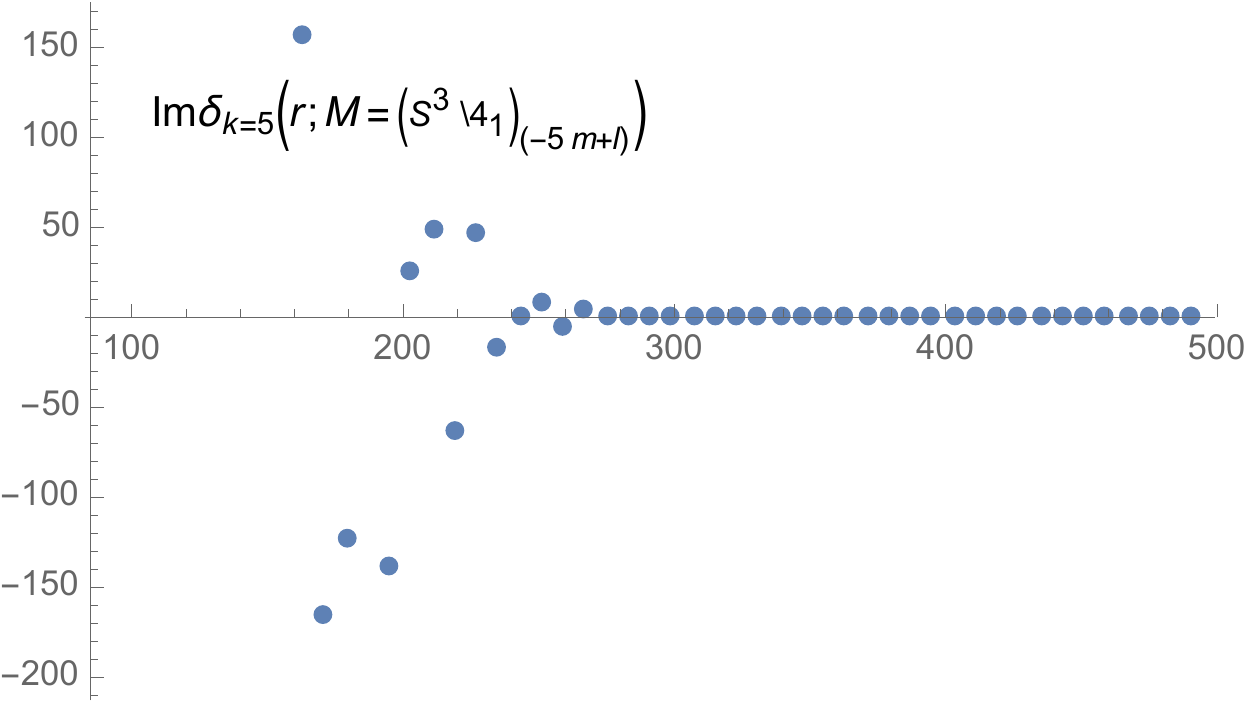}
   \includegraphics[width=.45\textwidth]{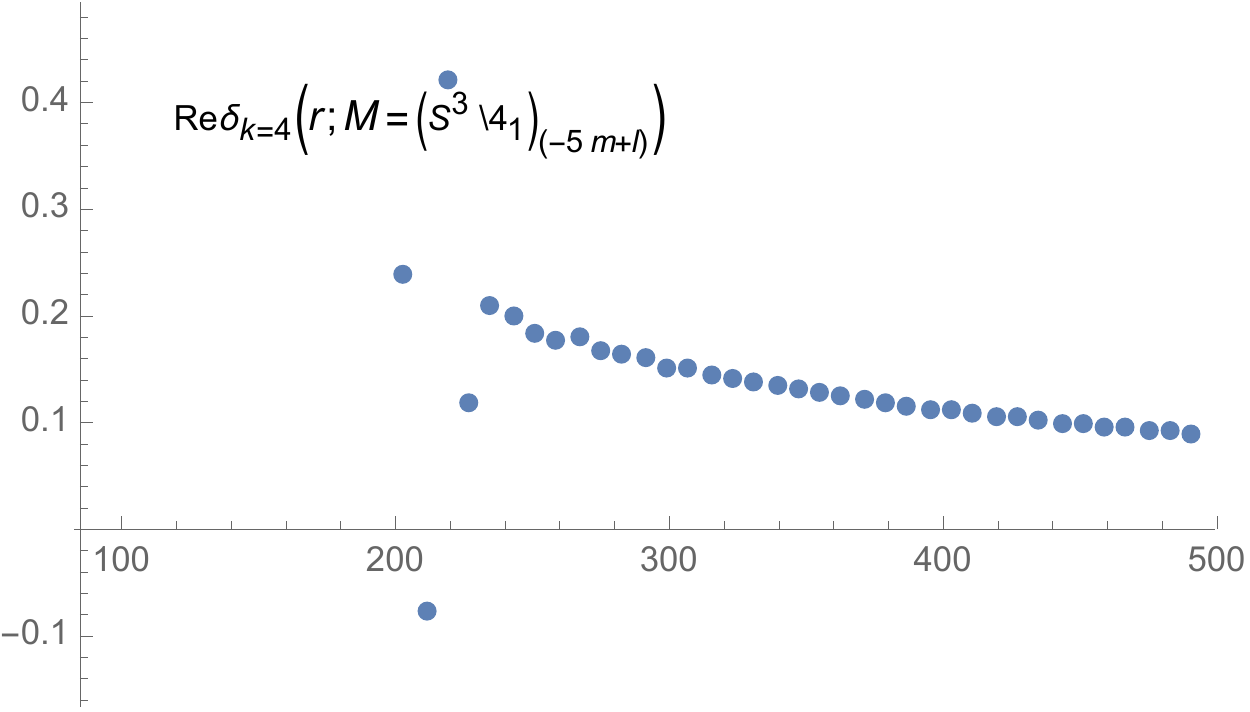}
   \includegraphics[width=.45\textwidth]{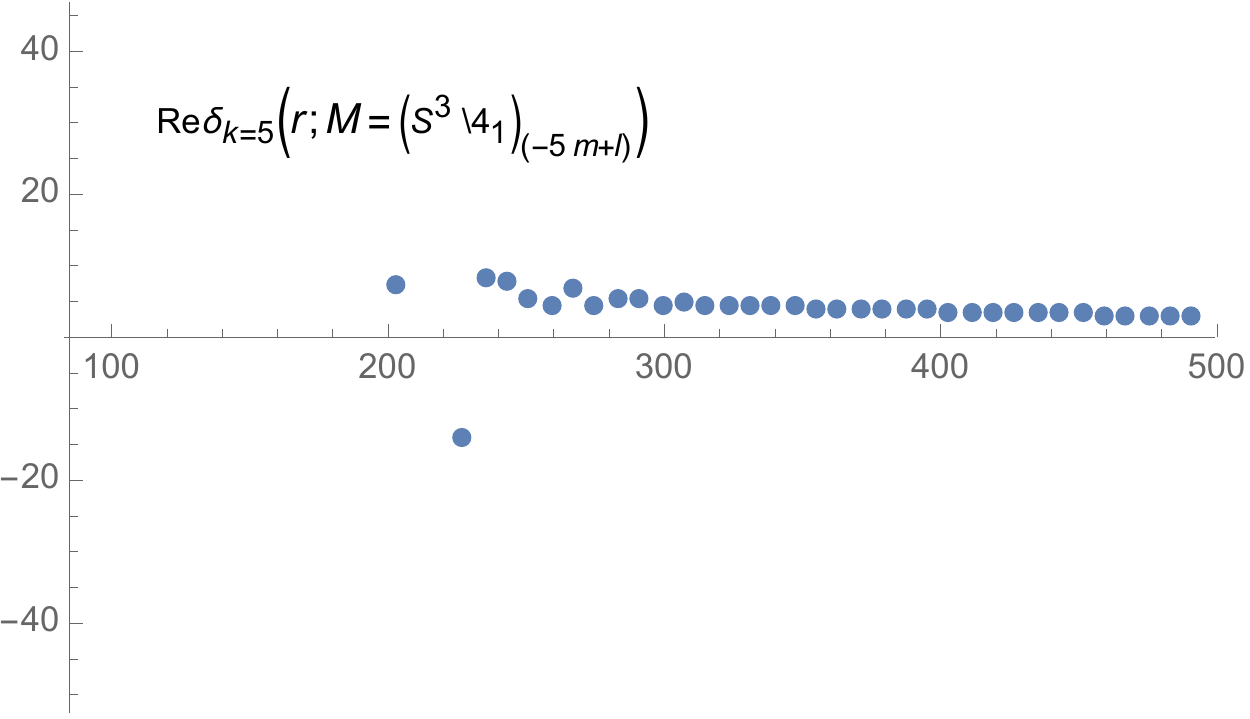}
   \end{center}
   \caption{Graphs of real and imaginary part of $\delta_k (r;M)$  for $k=4,5$ and $M=(S^3\setminus \mathbf{4}_1)_{-5 \sm+\mathfrak{l}}$.  The expression
   decreases quickly as $r$ becomes large, especially for larger $k$.
   }
    \label{fig:VCThurston}
\end{figure}

Let us take the example $M =(S^3\setminus \mathbf{4}_1)_{-5 \sm+\mathfrak{l}}=\textrm{(Thurston Manifold)}$.
 The colored Jones polynomial for the knot  $K=\mathbf{4}_1$ is (this is due to Habiro \cite{Habiro_some}, see also \cite{Masbaum_Habiro})
\begin{align}
J_N (\mathbf{4}_1;q) = \sum_{k=0}^{N-1}\prod_{i=1}^k (q^{N-i}-q^{-N+i})(q^{N+i}-q^{-N-i})\;.
\end{align}
Combining this expression with \eqref{so(3) WRT}, we can compute $\tau_r(M)$ for any $r$.
 Numerical value for the  perturbative expansions $S^{\rm hyp}_n(M)$ up to $n=5$ are given in \eqref{perturbative expansion for Thurston} (see also eq.~\eqref{relation between hyp and its conjugations}).
The numerical test for the generalized conjecture \eqref{Generalized VC for closed} is given in Fig.~\ref{fig:VCThurston}.
In these plots we defined
\begin{align}
\delta_k (r;M):=r^{k-1}  \big{[}\log \tau^{SO(3)}_{r}(M) - \sum_{n=0}^k \hbar^{n-1} S^{\rm hyp}_n(M) \big{]}_{\hbar = \frac{2\pi i}{r}} \;,
\end{align}
We see that both of $\textrm{Re}\, \delta_k (r;M)$ and $\textrm{Im}\, \delta_k (r;M)$ quickly decreases
as we increase the value of $r$.\footnote{For the imaginary part, we numerically fixed the phase ambiguity \eqref{ambiguity}; once we fix the integers
in \eqref{ambiguity} the rest can be used to the numerical values for $\textrm{Im}\, \delta_k (r;M)$.
} This is a highly non-trivial evidence for our conjecture.

\begin{figure}[htbp]
\begin{center}
   \includegraphics[width=.45\textwidth]{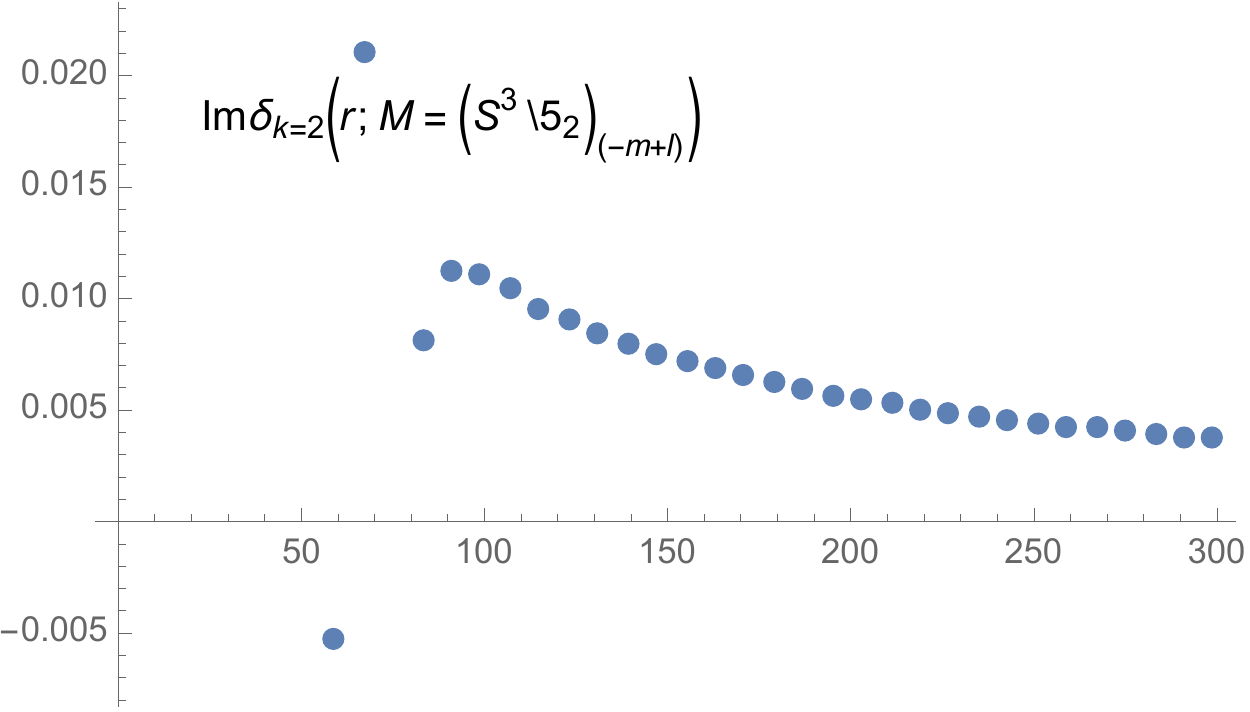}
     \includegraphics[width=.45\textwidth]{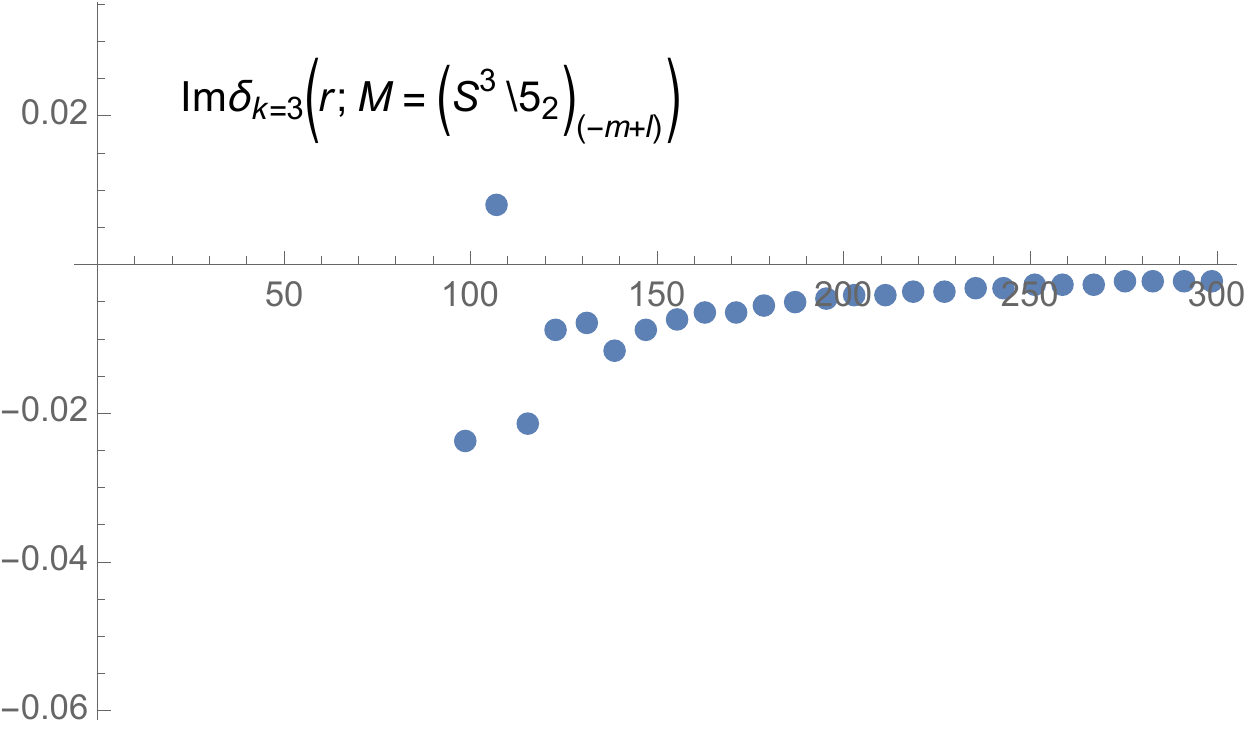}
     \\
   \includegraphics[width=.45\textwidth]{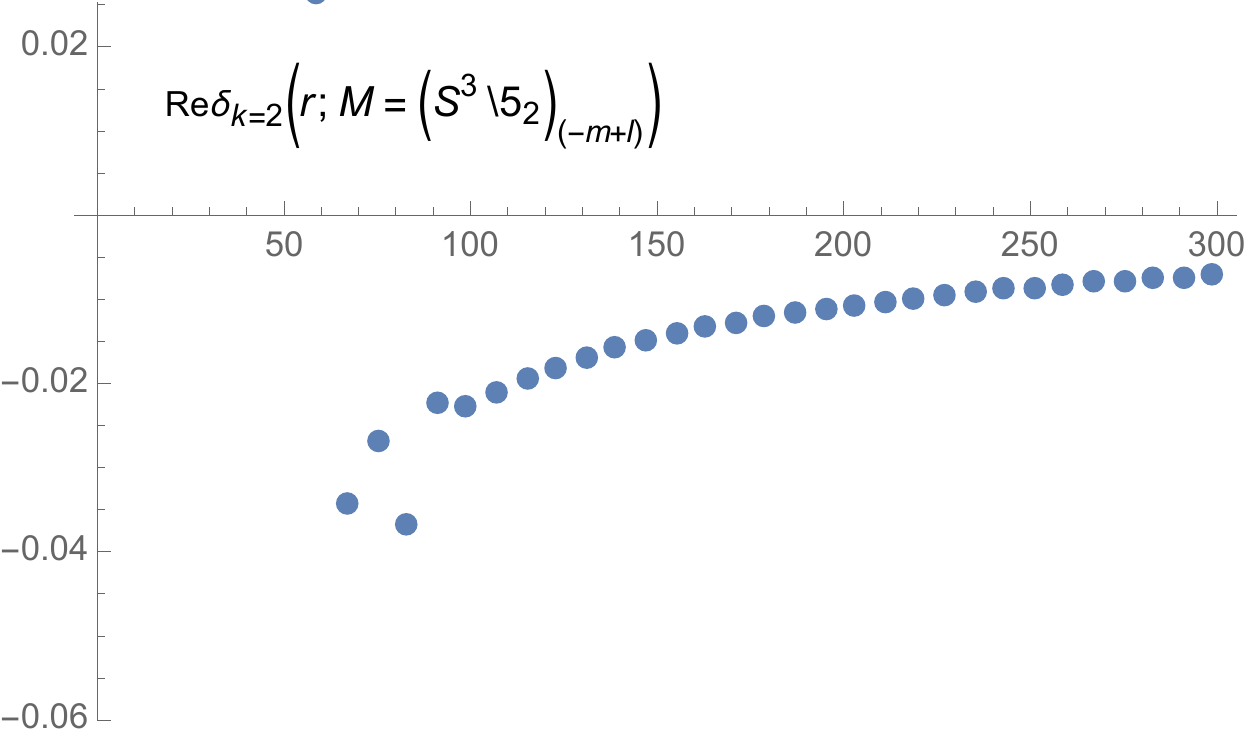}
   \includegraphics[width=.45\textwidth]{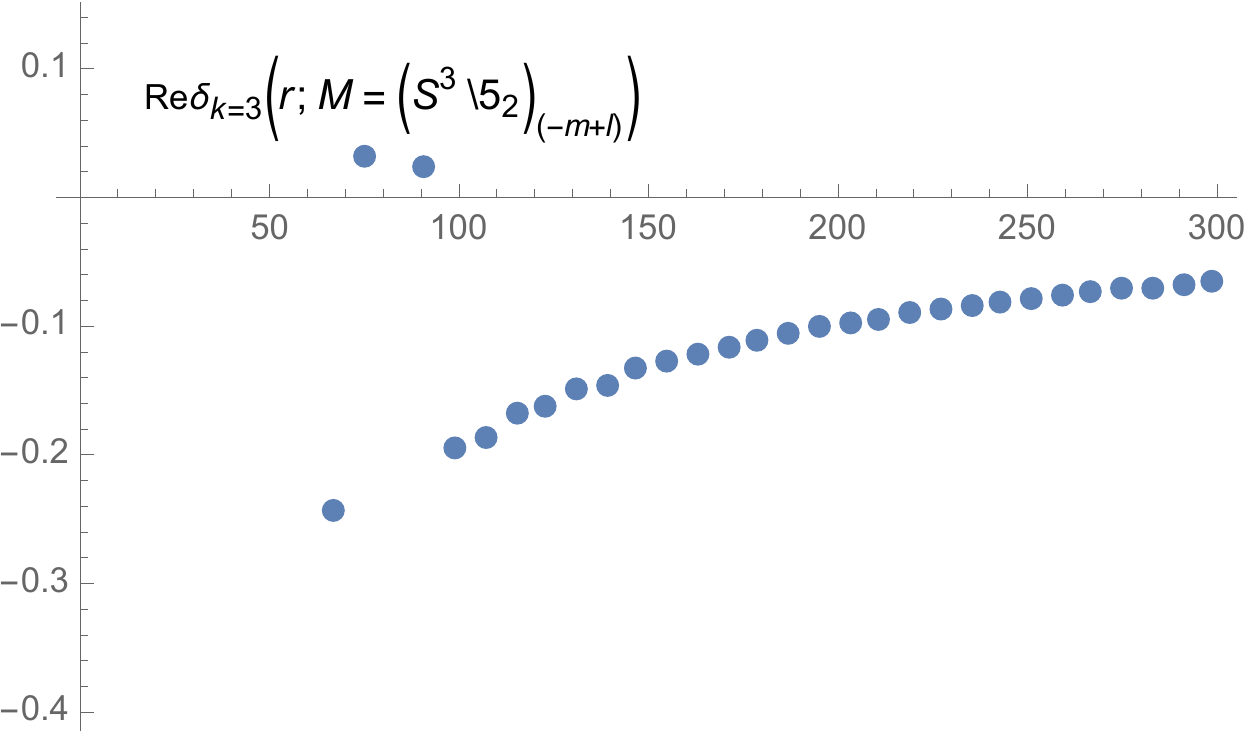}
   \end{center}
   \caption{Graphs of real and imaginary part of  $\delta_k \big{(}r;M=(S^3\setminus \mathbf{5}_2)_{- \sm+\mathfrak{l}} \big{)}$ for $k=2,3$.
   }
    \label{fig:VC2}
\end{figure}

As another example, let us consider a Dehn filling of the knot
$K=\mathbf{5}_2$. Its
colored Jones polynomial is \cite{Masbaum_Habiro}
\begin{align}
\begin{split}
&J_N (\mathbf{5}_2;q) = \sum_{k=0}^{N-1}c_k\prod_{i=1}^k (q^{N-i}-q^{-N+i})(q^{N+i}-q^{-N-i})\;,
\\
& c_k := (-1)^{k} q^{3k^2+5k} \sum_{i=0}^k q^{i^2-2i-3ki} \frac{[k]!}{[i]! [k-i]!}\;, \quad
[n]! := \prod_{a=1}^n \frac{q^{a}-q^{-a}}{q-q^{-1}}\;.
\end{split}
\end{align}
For a  manifold $M =(S^3\setminus \mathbf{5}_2)_{- \sm+\mathfrak{l}}=(S^3\setminus \mathbf{4}_1)_{- \sm+2\mathfrak{l}}$\footnote{The closed 3-manifold can be obtained by Dehn surgery  on $\mathbf{5}^2_1$:=(Whitehead link), $M=(S^3\backslash \mathbf{5}^2_1)_{\sm_1+\mathfrak{l}_1,-\sm_2+2\mathfrak{l}_2}$. From surgery calculus \cite{KirbyCalculus,rolfsen1984}, on the other hand, $(S^3\backslash \mathbf{5}^2_1)_{\sm_1+\mathfrak{l}_1}=(S^3\backslash \mathbf{4}_1) $ and $(S^3\backslash \mathbf{5}^2_1)_{-\sm_2+2\mathfrak{l}_2}=(S^3\backslash \overline{\mathbf{5}_2})$, where $\overline{K}$ means the mirror of a knot $K$. Thus $M= (S^3\backslash \mathbf{4}_1)_{-\sm+2\mathfrak{l}} =(S^3\backslash \overline{\mathbf{5}_2})_{\sm+\mathfrak{l}} =(S^3\backslash \mathbf{5}_2)_{-\sm+\mathfrak{l}} $.},
numerical value for the  perturbative expansions $S^{\rm hyp}_n(M)$ up to $n=3$ are given in \eqref{perturbative expansion for (4-1)(-1,2)} (see also eq.~\eqref{relation between hyp and its conjugations}). The numerical test for the generalized conjecture \eqref{Generalized VC for closed} is given in Fig.~\ref{fig:VC2}.

\section*{Acknowledgements}

We would like to thank Jun Murakami and Pavel Putrov for enlightening discussions, and the audience of the workshop ``mathematics and superstring theory'' (Kavli IPMU) for their feedback.
MY would like to thank Harvard university for hospitality where part of this work has been performed.
DG and MY is supported by WPI program (MEXT, Japan) and by JSPS-NRF Joint Research Project.
MY is also supported by JSPS Program for Advancing Strategic International Networks to Accelerate the Circulation of Talented Researchers, by JSPS KAKENHI Grant No. 15K17634. MR gratefully acknowledges the support of the Institute for Advanced Study, DOE grant DE-SC0009988 and the Adler Family Fund.

\appendix
\section{Review: Surgery Construction of Closed 3-manifolds }

In this appendix we quickly summarize the concept of
Dehn filling, for readers unfamiliar with the concept.

Suppose that $K$ is a knot inside a closed 3-manifold $\hat{M}$, namely has only one component,
and consider a knot complement
\begin{align}
\hat{M}\setminus K \;. \label{knot complement}
\end{align}
The boundary of this 3-manifold is a two-torus $T^2$,  which has two non-trivial one-cycles, $\mathbb{X}$ and $\mathbb{P}$
\begin{align}
\langle \mathbb{X}, \mathbb{P}\rangle = H_1 \big{(}T^2 = \partial (\hat{M}\backslash K)  , \mathbb{Z} \big{)} = \mathbb{Z} \oplus \mathbb{Z}\;.
\label{XP_def}
\end{align}
Now a $(p,q)$-Dehn filling of $M$, which we denote as $(\hat{M}\backslash K)_{p\mathbb{X}+q\mathbb{P}}$, is obtain by
eliminating the boundary torus of the knot complement by
filling in solid tori $D^2\times S^1$,
so that the combination $p \mathbb{X}+ q \mathbb{P}$ is
contractible inside $D^2\times S^1$ :
\begin{align}
(\hat{M}\backslash K)_{p\mathbb{X}+q \mathbb{P}}:=  (D^2\times S^1)  \bigcup_{\varphi\in \mathop{\mathrm{Aut}}(T^2)}(\hat{M}\backslash K) \;.
\label{Dehn_filling}
\end{align}
where an automorphisms $\varphi$ on the two-torus
is taken to be\footnote{The topological type of the resulting 3-manifold
is invariant under continuous deformation of $\varphi$ and hence we can regard
$\varphi$ as an element of the $PSL(2, \mathbb{Z})$.}
\begin{align}
\begin{split}
\varphi &: H_1 (\partial(\hat{M}\backslash K) , \mathbb{Z})  \rightarrow H_1 (\partial(D^2\times S^1) , \mathbb{Z})  \;,
\\
& \left(\begin{array}{c} \mathbb{X} \\ \mathbb{P}\end{array}\right) \rightarrow \; g \cdot \left(\begin{array}{c}\a \\ \b \end{array}\right), \;\textrm{with }    g:= \pm \left(
\begin{array}{cc}
r & s \\
p& q
\end{array}
\right)
\in PSL(2, \mathbb{Z}) \;.\label{varphi}
\end{split}
\end{align}
Here $(\alpha, \beta)$ is a basis  of $H_1\big{(}\partial (D^2\times S^1),\mathbb{Z}\big{)}$ :
\begin{align}
\a \sim (S^1 \textrm{ in }D^2\times S^1)\;, \quad \beta \sim (S^1 \textrm{ of } \partial D^2 )\;.
\end{align}
The definition \eqref{Dehn_filling} apparently depends on the
choice of integers $(r,s)$ in \eqref{varphi}, in addition to $(p,q)$.
Indeed, given $(p,q)$ there is an ambiguity in the choice of $(r,s)$ as
$(r,s)\to (r,s)+n(p,q)$, with $n\in \mathbb{Z}$.
However, this ambiguity is equivalent with the ambiguity of the
longitude inside a general 3-manifold, and keeps the
topology of the manifold after the Dehn filling (see e.g.\ discussion around Figure 7 of \cite{Gang:2015wya}).
Consequently
this ambiguity preserves the partition function of the complex Chern-Simons theory,
possible up to some overall pre-factors originating from framing anomaly.
Note also that we also have an overall sign ambiguity $(p, q)\sim (-p, -q)$, which we can eliminate by considering a slope $p/q$.

In general, the link $L$ has several (say $S$) connected components and we can
choose $(p_{\alpha} ,q_{\alpha})$ Dehn-surgeries for the $\alpha$-th component,
for some value of $\alpha$ ($1\le \alpha\le S$). The resulting 3-manifold is then a complement of a link with $S-1$ components. When do the Dehn filing on all the link complements we obtain a closed 3-manifold $(\hat{M} \setminus  L)_{p_1 \mathbb{X}_1 +q_1 \mathbb{P},\ldots , p_S \mathbb{X}_S+q_S \mathbb{P}_S}$.

\section{Review: State-Integral Model for Link Complement}\label{sec.state_integral}

Let us first review the state-integral model for a knot/link complement $\hat{M}\setminus L$, following \cite{Dimofte:2011gm,Dimofte:2012qj}.
We consider a regular ideal triangulation of  a link complement $\hat{M}\setminus L$:
\begin{align}\label{eq.triangulation}
\hat{M}\setminus L = \left( \bigcup_{i=1}^k \Delta_i \right)/\sim\;.
\end{align}
where each $\Delta_i$ is an ideal tetrahedron (ideal here means that all the vertices are located on the boundary).
The symbol $\sim$ denotes the gluing of the $k$ tetrahedra.
The shape of an ideal tetrahedron can be parametrized by a shape parameter $z_i$.
This is a complexification of the dihedral angles between two faces, and once we fix the choice of an edge the remaining dihedral angles are given by
\begin{align}
z_i'=(1-z_i)^{-1}, \quad z_i''=1-z_i^{-1}, \quad z_i z'_i z''_i=1\;.
\label{zzbar}
\end{align}
In this parametrization, we fixed a choice of which dihedral angle to call $z_i$ (and not $z'_i$ or $z''_i$). Such a choice is called the `quad type'.

We next impose extra conditions originating from the gluing of tetrahedra.
First we have gluing conditions at each internal edge.
It follows from vanishing of the Euler number of the boundary tori
that the number of edges is also given by $k$.
We also need to impose conditions on the cups boundaries, and we therefore have extra $S$ conditions. This naively means that we have $k+S$ conditions.
However, it turns out that only $k-S$ out of the $k$ conditions from internal edges
are linearly independent, leading to total of $(k-S)+S=k$ constraints.

To describe this gluing, let us denote by $z_i$ the shape parameter (modulus) of the $i$-th
ideal tetrahedron ($i=1, \ldots, k$). We can then express the constraint equation as \cite{NeumannZagier}
\begin{align}
\prod_{j=1}^k z_j^{A_{ij}} z''_j{}^{B_{ij}}=(-1)^{ \nu_i} \;, \quad i=1\;,\ldots,\;k\;,
\label{eq.gluing_1}
\end{align}
where the matrices $A=(A_{ij}), B=(B_{ij})$ are $k\times k$-matrices with integer entries, and
$\vec{\nu}$ is an $k$-component integer vector.   When we consider deformations of the boundary holonomy, the
equations are modified to be
\begin{align}
\prod_{j=1}^k z_j^{A_{ij}} z''_j{}^{B_{ij}}=e^{2X_i+i\pi \nu_i} \;,\quad i=1\;,\ldots,\;k\;.
\label{eq.gluing_2}
\end{align}
Here a length-$k$ vector $\vec{X}=(X_i)$ is defined by
from a set of $S$ parameters $(X_{\alpha})_{\alpha=1}^S$ to be
\begin{align}
&X_{i} := \begin{cases}  X_i &\mbox{if } 1\leq i \leq S  \\
0 & \mbox{if } S+1 \leq i \leq k \end{cases}\;,
\label{X_def}
\end{align}
where we have chosen the indices $I$ such that
the first $S$ conditions ($i=1, \ldots, S$) come from cusp boundaries,
and the remaining $k-S$ conditions ($i=S+1\ldots, k$) from internal edges.
The parameters $(X_\a )_{\a=1}^S$ (or rather its exponential, to match with the standard definition in literature) parameterize the boundary $PSL(2,\mathbb{C})$-holonomies  along  $S$ one-cycles $\mathbb{X}_\a$.
\begin{align}
P \exp \left(\oint_{\mathbb{X}_\a} \mathcal{A}\right) \sim \pm \left(\begin{array}{cc} e^{X_\a} & * \\0 & e^{-X_\a}\end{array}\right)\;
\end{align}
with a conjugacy equivalence  relation $\sim$. Similarly, we can introduce boundary holonomy variables $P_\a$ along $S$ $\mathbb{P}_\a$ and these variables also can be written in terms of  shape parameters
\begin{align}
\prod_{j=1}^k z_j^{\hat{C}_{\a j} } z_j''^{\hat{D}_{\a j}} = e^{P_\a +i \pi \hat{\nu}'_\a}\;, \quad \a=1\;,\ldots,\;S\;.
\end{align}
In general, the $\hat{C},\hat{D}$ and $\hat{\nu}'$ are valued in half-integers. The partition function for the state-integral model is given by \eqref{Dimofte integral}
in the main text.
As discussed in section \ref{sec.volume}, in the semiclassical limit, its saddle point value reproduces the shape modulus of the tetrahedron by the relation $e^{Z_i}=z_i$.
The matrices ($\hat{C},\hat{D}$) can be extended to $k\times k$ matrices $(C,D)$ in a way that \cite{NeumannZagier}.
\begin{align}
\begin{split}
& C_{i j} = \hat{C}_{ij}\;, \; D_{ij} = \hat{D}_{ij} \quad \textrm{for $i=1,\ldots, S$} \;,
\\
&\textrm{and }\left(\begin{array}{cc}A & B \\ C & D\end{array}\right)\in \mathrm{Sp}(2k, \mathbb{Q}) \;.
\label{Symplectic property}
\end{split}
\end{align}
Similarly, the vector $\hat{\nu}'$ is extended to $k$-component vector $\vec{\nu}'$:
\begin{align}
\vec{\nu}' := (\hat{\nu}_1,\ldots, \hat{\nu}_S,0,\ldots,0)\;.
\end{align}
The vectors $(\vec{f}, \vec{f''})=(f_i, f''_i)_{i=1}^k$
are known as combinatorial flattening, and satisfy the constraints
\begin{align}
&\vec{\nu}=A \cdot \vec{f}+  B \cdot \vec{f}''  \;, \quad \vec{\nu}'=C \cdot \vec{f}+  D \cdot \vec{f}'' \;.
\label{nu_def}
\end{align}

\bibliographystyle{nb}
\bibliography{w3c}

\end{document}